\documentclass[letterpaper,11pt]{article}
\pdfoutput=1

\usepackage{jheppub}
\usepackage{multirow}

\usepackage{subcaption}
\usepackage[countmax]{subfloat}
\usepackage{amssymb}
\usepackage{amsmath}
\usepackage{color}
\usepackage{graphicx}
\usepackage{verbatim}
\usepackage{amsthm}
\usepackage{slashed}
\usepackage{hyperref}
\usepackage{makecell}

\preprint{}

\title{
Constraints on neutrino non-standard interactions from LHC data with large
missing transverse momentum
}

\author{DianYu Liu$^1$, ChuanLe Sun$^1$, Jun Gao$^{1,2}$}

\affiliation{$^1$INPAC, Shanghai Key Laboratory for Particle Physics and Cosmology, School of
Physics and Astronomy, Shanghai Jiao-Tong University, Shanghai 200240, China}
\affiliation{$^2$Center for High Energy Physics, Peking University, Beijing 100871, China}

\emailAdd{dianyu.liu@sjtu.edu.cn}
\emailAdd{chlsun60@sjtu.edu.cn}
\emailAdd{jung49@sjtu.edu.cn}

\abstract{
The possible non-standard interactions (NSIs) of neutrinos with
matter plays important role in the global determination of neutrino
properties. 
In our study we select various data sets from LHC measurements
at 13 TeV with integrated luminosities of $35\sim 139$ fb$^{-1}$,
including production of a single jet, photon, $W/Z$ boson, or charged lepton
accompanied with large missing transverse momentum.
We derive constraints on neutral-current NSIs with quarks imposed by different
data sets in a framework of either effective operators or simplified
$Z'$ models.
We use theoretical predictions of productions induced by NSIs
at next-to-leading order in QCD matched with parton showering 
which stabilize the theory predictions and result in more robust constraints.
In a simplified $Z'$ model we obtain
a 95\% CLs upper limit on the conventional NSI strength $\epsilon$ of 0.042 and
0.0028 for a $Z'$ mass of 0.2 and 2 TeV respectively.
We also discuss possible improvements from future runs of LHC
with higher luminosities.
}
  

\begin{document}

\maketitle

\section{Introduction}

Confirmation on neutrino oscillation in recent decades requires
non-vanishing neutrino masses. 
The effective operator with lowest dimensions that respects Standard
Model(SM) gauge symmetry is the dimension-five Weinberg operator.
This operator can give rise to neutrino masses, and can be
achieved by different UV-complete models depending on the portal
particle~\cite{Minkowski:1977sc,Mohapatra:1980yp,Yanagida:1980xy,Foot:1988aq}.
Many extensions of the SM such as Supersymmetry also introduce
dimension-six operators of the form
\begin{equation}\label{eq:1}
  \mathcal{L}_{NSI,CC} = -2\sqrt{2} G_F \epsilon_{\alpha \beta}^{ff',Y}
(\bar{\nu}_{\alpha} \gamma_{\mu} P_L l_{\beta} )(\bar{f'} \gamma^{\mu} P_Y f) + h.c.
\end{equation}
known as Charged-Current Non-Standard neutrino Interaction(CC NSI) and
\begin{equation}\label{eq:2}
  \mathcal{L}_{NSI,NC} = -2\sqrt{2} G_F \epsilon_{\alpha \beta}^{f,Y}
	(\bar{\nu}_{\alpha} \gamma_{\mu} P_L \nu_{\beta} )(\bar{f} \gamma^{\mu} P_Y f) + h.c.
\end{equation}
known as Neutral-Current NSI (NC NSI), that was first proposed
by Wolfenstein in 1977 to explain the neutrino
oscillation~\cite{Wolfenstein:1977ue}. 
$\epsilon_{\alpha \beta}^{ff',Y}$ and $\epsilon_{\alpha \beta}^{f,Y}$
define the strength of NSIs respectively, $\alpha, \beta \in \{e,\mu,\tau\}$, $f$
denote charged leptons or quarks and $P_Y$ is chiral projection operator($P_L$ or $P_R$).
CC NSI modifies the neutrino production and detection through
its effect on processes such as muon decay and inverse beta
decay~\cite{C.:2019dbf,Santos:2020dgs,Yue:2020wkj}.
From those processes severe bound can be obtained~\cite{Biggio:2009nt}.
NC NSI plays an important role in neutrino oscillation experiments
due to modification to the effective Hamiltonian, especially
the matter potential~\cite{Wolfenstein:1977ue,Mikheev:1986gs,Ohlsson:2012kf,Chatterjee:2018dyd,Ge:2018uhz,Babu:2019iml,Bakhti:2020hbz,Garcia:2020jwr}. 
This modification further leads to nuisances and degeneracies to the
measurement of neutrino oscillation parameters~\cite{Fornengo:2001pm,Miranda:2004nb,Liao:2016hsa,Coloma:2016gei,Esteban:2018ppq,Esteban:2019lfo,Dutta:2020che,Esteban:2020itz}. 
On the contrary to the tight bounds on CC NSI, NC NSI is
less constrained and has been studied extensively~\cite{Davidson:2003ha,Biggio:2009nt}.
There are many experiments giving constraints on NSIs such as
XENON1T~\cite{Karmakar:2020rbi}, KM3NeT-ORCA~\cite{KhanChowdhury:2020qqu}, 
IceCube~\cite{Esmaili:2013fva,Esmaili:2018qzu}, DUNE~\cite{Masud:2015xva,deGouvea:2015ndi,Verma:2018gwi,Masud:2018pig,Giarnetti:2020bmf}, 
Super-Kamiokande~\cite{Mitsuka:2011ty} and Borexino Phase II~\cite{Agarwalla:2019smc}.

Long-baseline(LBL) experiments are the next generation neutrino oscillation
experiments for their sensitivity to neutrino mass ordering and CP
violating phase $\delta_{CP}$~\cite{Ribeiro:2007ud,Miranda:2015dra,Flores:2018kwk,
Feng:2019mno,Yasuda:2020cff,Denton:2020uda}. 
In recent T2K measurement, a preference of normal mass ordering and a best-fit
value of $\delta_{CP}$ at -1.89 corresponding to this ordering is
reported~\cite{Abe:2019vii}. 
The incorporation of NC NSI however complicates the analysis by
introducing new CP violating sources and parameter
degeneracies~\cite{Chatterjee:2020kkm}. 
The degenerate LMA-Dark solution results in a preference for the inverted mass
ordering and therefore an almost total loss of sensitivity thereof~\cite{Esteban:2019lfo}.

Compared with oscillation experiments, neutrino scattering experiments are
complementary for two reasons. 
First, the parameter degeneracies are broken-down since for scattering
experiments, the measured cross section subjects to no periodicity and
there is no unobservable overall phase factor from wave functions. 
This further makes it possible to constrain individual
diagonal parameters in the effective Hamiltonian rather than their differences.
Second, measurements in oscillation experiments generally depend on the
composition of the media, while scattering experiments are less flavor dependent.
CHARM and NuTeV experiments report the ratios of neutral-current and
charged-current neutrino-nucleon deep-inelastic scattering cross
sections~\cite{Dorenbosch:1986tb,Zeller:2001hh}.
In the presence of NSI, the ratios of cross sections are modified and
are constrained by experimental measurements. 
For NSI induced by heavy mediator with mass larger than 
the experimental energy scale, 
bounds on NSI parameters ranging from sub-percent level to a few percent level 
are obtained 
by a global fit to data from current oscillation experiments and 
the two scattering experiments~\cite{Coloma:2017egw}, under assumption that NSI affects only up or down quark at a time. 
Strong constraints on NSI parameters involving $\mu$ and $\tau$ flavors are also obtained. 
For mediator with mass lower than $\mathcal{O}$(GeV), 
the contact-interaction approximation is invalid 
in deep-inelastic scattering energy range, 
but yet still work in coherent neutrino-nucleus scattering(CE$\nu$NS) 
of which the momentum transfer lies at $\mathcal{O}$(10 MeV). 
In this scenario, similar bounds can be set taking advantage of the recent COHERENT measurement~\cite{Akimov:2017ade,Liao:2017uzy,Denton:2018xmq,Billard:2018jnl,Altmannshofer:2018xyo,Miranda:2019skf,Giunti:2019xpr,Coloma:2019mbs,Canas:2019fjw,Han:2019zkz}. 
These works also make it clear that in combination with data from
scattering experiments degeneracies on neutrino parameters
can be resolved to some extent.
High-energy colliders can also help with study of NSIs. 
In previous works, limits on the NSIs from $e^{+} e^{-}$ colliders and the 
LHC are given by~\cite{Berezhiani:2001rs,Barranco:2007ej} and~\cite{Friedland:2011za,Choudhury:2018azm,Babu:2020nna} respectively. 
Other new physics searches such as Dark Matter, Supersymmetry have also been studied at LHC~\cite{He:2007ge,Chao:2008mq,Cao:2009uw,Bai:2012xg,Bell:2012rg,Zhou:2013fla,Lin:2013sca,Mao:2014rga,Bell:2015sza,Abercrombie:2015wmb,Neubert:2015fka,Sirunyan:2017cwe,Aguilar-Saavedra:2019iil}.
LHC offers a unique way to study neutrino physics for neutrino energy larger
than 300 GeV~\cite{Beni:2020yfy}. 
Different from oscillation and other scattering experiments, the flavor of NSI is indistinguishable at LHC. 
Besides, LHC is sensitive to both vector-like and axial vector like NSI as opposite to oscillation experiments that only the former relates. 
Thus the LHC experiment plays a further complementary role 
in searches of NSI~\cite{Davidson:2011kr,Dev:2019anc,Babu:2019mfe}. 
Neutrinos produced by NSI at the LHC would leave large unbalanced 
transverse energy or momentum in detectors. 
The major irreducible SM backgrounds are from 
the decay of $W$ and $Z$ bosons to neutrinos. 
Meanwhile, an underlying theory model, however, is generally needed since in this scenario 
the validity of effective field theory (EFT) approach is no longer guaranteed. 
Simplified $Z'$ models with possible UV-completions have
been considered~\cite{Bell:2015rdw,Farzan:2017xzy,Heeck:2018nzc,Pandey:2019apj,Farzan:2019xor,Flores:2020lji,Alvarez:2020yim}. 
And given that the mass of $Z'$ boson is much larger than momentum
transfer at LHC, one can come back to the EFT case. 
It is noted that similar signals can be produced for various models with dark matters
at the LHC. 
To discriminate these two sources, one can add the shape of distribution of missing
energy to the data analysis~\cite{Franzosi:2015wha}. 
Also, given consideration that neutrinos are produced along with charged leptons due
to the $SU(2)_L$ doublet nature, data from multi-lepton channel can be complemented
to give further discrimination~\cite{Franzosi:2015wha,Han:2019zkz}.

In this paper, we focus on the aforementioned $Z'$ model and study constraints
on NC NSI parameters based on various measurements at LHC 13 TeV with large missing
transverse momentums in the final states. 
We considered data sets on production of mono-jet, mono-$W/Z$ boson, mono-lepton and
mono-photon recorded by both ATLAS and CMS collaborations. 
We conclude the CMS mono-jet data imposes the strongest constraints,
and a flavor-blind bound of a few per mille has been obtained for $Z'$
mass around 2 TeV. 

The rest of this paper is organized as follows. 
In Section \ref{sec:aa}, we discuss the model assumptions and the resultant constraints. 
In Section \ref{sec:mass}, we discuss the theoretical uncertainties. 
The LHC combination and projections are presented in Section \ref{sec:thunc}, and we conclude
in Section \ref{sec:conc}.

\section{Model assumptions and constraints}\label{sec:aa}

Phenomenologies of neutrino non-standard interactions can span an
energy range from MeV in neutrino oscillations to TeV at high energy
colliders, or even higher in reaction of cosmic neutrinos.
Simplified models or descriptions of NSIs are always adopted in various
analyses and for easy comparison of constraints from different experiments.
We outline the simplified models used in our study and then the
constraints obtained with LHC data.

\subsection{Theoretical setup}

It is justified to express the NSI in a model-independent manner
using the effective field theory framework in neutrino study at low energies,
for example in study of neutrino oscillations.  
NC NSI between neutrino and matters can be described 
by dimension-six four fermion operators in the EFT framework
as\cite{Wolfenstein:1977ue,Gavela:2008ra,Han:2020pff}
\begin{equation}
\mathcal{L}_{\mathrm{NSI}}=-2 \sqrt{2} G_{F} \epsilon_{\alpha \beta}^{f, Y}
	\left(\bar{\nu}^{\alpha} \gamma_{\mu} P_{L} \nu^{\beta}\right)
\left(\bar{f} \gamma^{\mu} P_{Y} f\right),
\label{EFT_NSI}
\end{equation}
where $G_{F}$ is the Fermi constant, $\epsilon_{\alpha\beta}$ is
the strength of NSIs,
$\alpha, \beta$ denotes the lepton flavors $\{e, \mu, \tau\}$,
and $f$ can be either charged leptons or quarks.
$P_Y$ can be $P_{L}$ or $P_{R}$ which are chiral projectors of left-handed and
right-handed.
In our study we focus on NSI between neutrinos and quarks of
both up and down-type $f=\{u,\, d\}$.
In general the above operators can be embedded into a gauge invariant
operator from integrating out heavy degree of freedoms of new physics,
\begin{equation}
-\frac{c}{\Lambda^{2}}\left(\bar{L}_{\alpha} \gamma_{\mu} L_{\beta}
	\right)\left(\bar{Q}_Y \gamma^{\mu} Q_Y\right),
\label{EFToperator}
\end{equation}
where $L$ is the $SU(2)_L$ doublet of leptons, $Q_Y=\{Q_L, u_R, d_R\}$ are
$SU(2)_L$ doublet or singlet of quarks. 
$\Lambda$ is the typical scale of the new physics models
and $c$ is the Wilson coefficient.
The conventional NSI strength $\epsilon$ can be related to $\Lambda$
as $\epsilon_{\alpha\beta}=c/(2\sqrt{2}G_{F}\Lambda^{2})$.
Stringent limits on NSI exist due to various measurements on charged
leptons at colliders once the interaction also involves
charged leptons as in Eq.~(\ref{EFToperator}), for example see Ref.~\cite{Han:2019zkz}
for recent discussions.
At high energies, for instance at the LHC, effects of neutrino
NSI may not be simply described by EFT operators since the
momentum transfers can be sufficiently high to resolve further dynamics 
of the new physics.
In this study we focus on a simplified model with NSI between neutrinos
and quarks induced by $s$-channel exchange of a $Z^{\prime}$ boson.
This simplified model is more appropriate than the aforementioned 
EFT description at high energy regions. 
The effective Lagrangian of the interactions can be written as~\cite{Babu:2020nna}
\begin{equation}
\mathcal{L}_{\mathrm{NSI}}^{Z'}=-\left(g_{\nu}^{\alpha \beta}
\bar{\nu}_{\alpha} \gamma^{\mu} P_{L} \nu_{\beta}+g_{q }^{Y} \bar{q}
	\gamma^{\mu} P_{Y} q\right) Z_{\mu}^{\prime},
\label{zp}
\end{equation}
where $Z^{\prime}_{\mu}$ represents the force mediator with
mass $M_{Z^{\prime}}$. 
We assume the interactions are independent of quark generations, and
only contain vector current for simplicity, namely $g_u^L=g_u^R\equiv g_u$
and $g_d^L=g_d^R\equiv g_d$.
At low energies or for a $Z'$ boson with sufficiently large mass
the $s$-channel $Z'$ model can be matched onto the
EFT representation defined in Eq.~(\ref{EFT_NSI}) with 
\begin{equation}
	\epsilon_{\alpha\beta}^{u(d),V} \equiv \frac{g_{\nu}^{\alpha \beta}
	g_{u(d)}}{2 \sqrt{2} G_{F} M_{Z^{\prime}}^{2}},
\label{conv}
\end{equation}
where the superscript $V$ indicates a vector-current form
on the matter side in Eq.~(\ref{EFT_NSI}). 
There are many new physics models on ultraviolet completion of neutrino
NSI such as Zee Model\cite{Zee:1980ai} and One-Loop LQ
Model\cite{AristizabalSierra:2007nf,Dorsner:2017wwn}.

\begin{figure}[tbp]
  \centering
  \subcaptionbox{mono-jet: $q \bar{q} \to g \nu \bar{\nu}$}[7.7cm] 
    {\includegraphics[width=7.7cm]{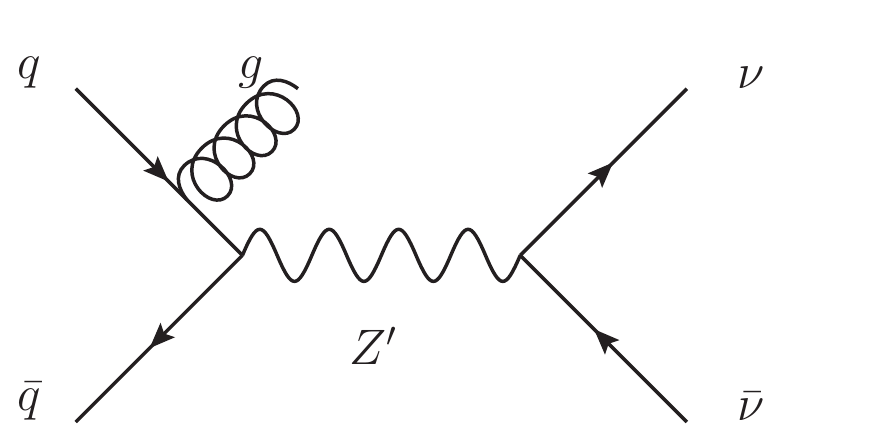}}
  \subcaptionbox{mono-photon: $q \bar{q} \to \gamma \nu \bar{\nu}$}[7.7cm]
    {\includegraphics[width=7.7cm]{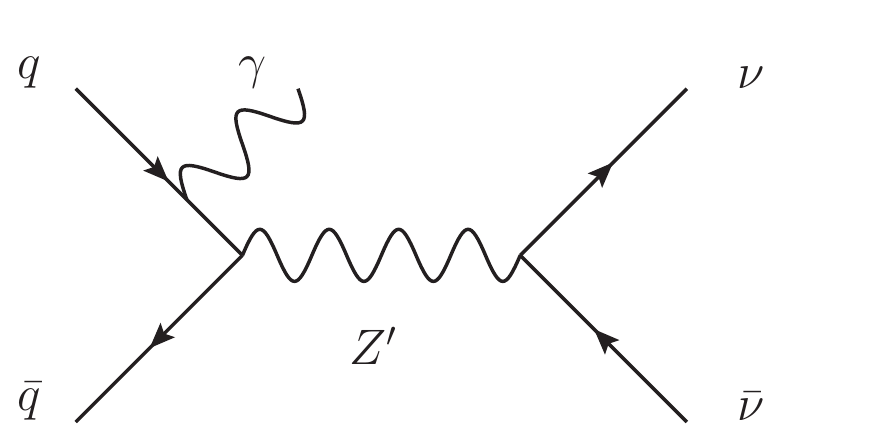}}
  \subcaptionbox{mono-lepton: $q \bar{q}^{\prime} \to l \bar{\nu} \nu \bar{\nu}$}[7.7cm]
    {\includegraphics[width=7.7cm]{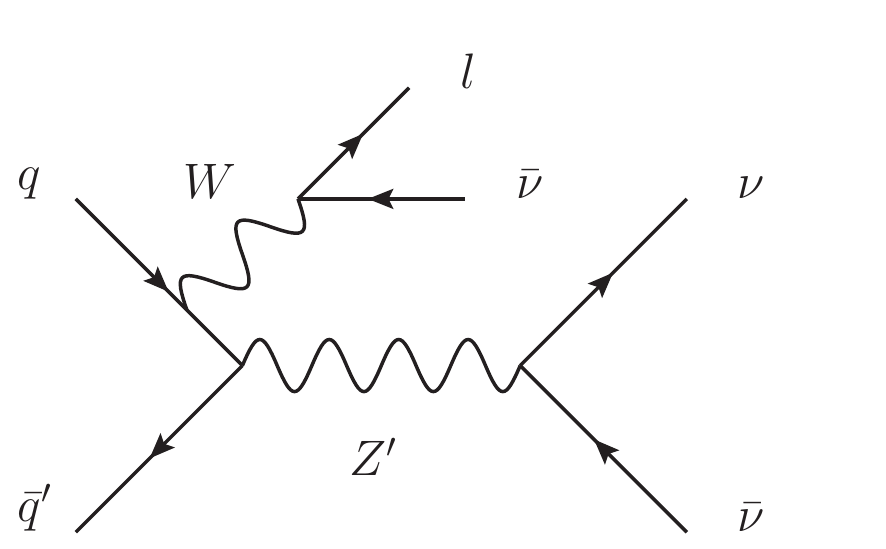}}
  \subcaptionbox{mono-W(Z): $q \bar{q}^{\prime} (\bar{q}) \to W (Z)  \nu \bar{\nu}$}[7.7cm]
    {\includegraphics[width=7.7cm]{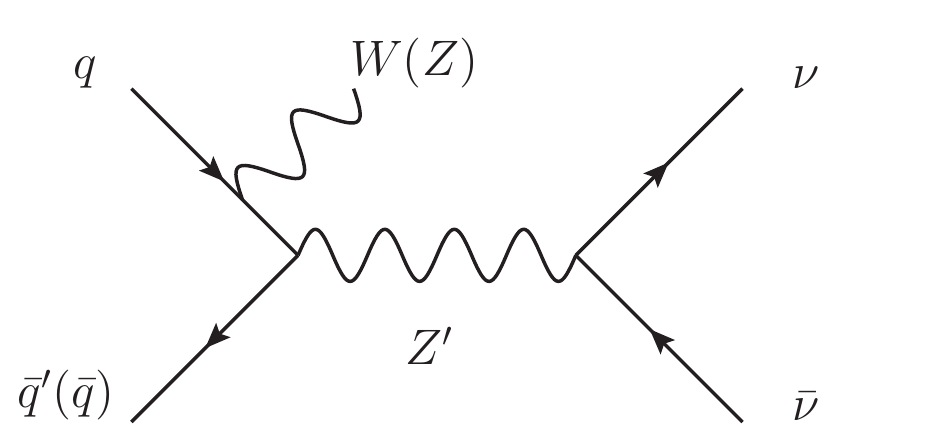}}
  \subcaptionbox{mono-Z: $q \bar{q} \to Z \nu \bar{\nu}$}[7.7cm]
    {\includegraphics[width=7.7cm]{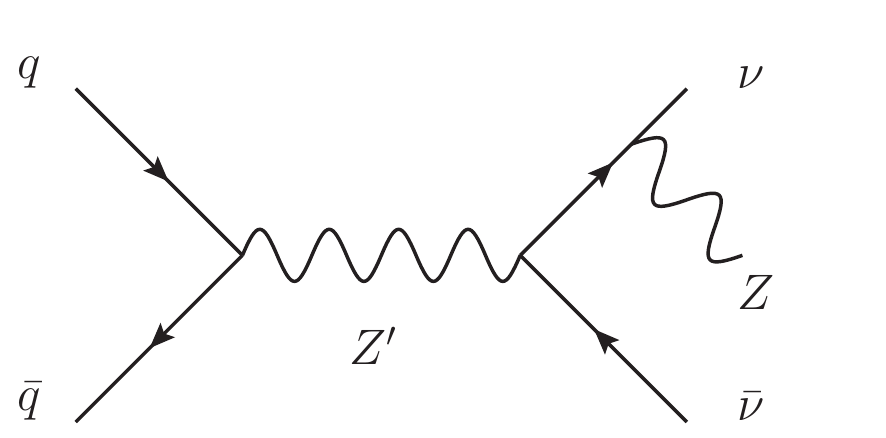}}
  \caption{Representative Feynman diagrams at leading order for the process
	$q \bar{q} \to g \nu \bar{\nu}$, $q \bar{q} \to \gamma \nu \bar{\nu}$,
	$q \bar{q}^{\prime} \to l \bar{\nu} \nu \bar{\nu}$, 
	$q \bar{q}^{\prime} (\bar{q}) \to W (Z)  \nu \bar{\nu}$
	and $q \bar{q} \to Z \nu \bar{\nu}$. Flavor indices of neutrinos
	are suppressed for simplicity.}
  \label{Fig:treelevel}
\end{figure}

In this study we utilize experimental measurements on signature with
large missing transverse momentum at the LHC to constrain
neutrino NSI. 
We select recent ATLAS and CMS data sets on production of mono-jet,
mono-photon, mono-$W/Z$ and mono-lepton.
The representative Feynman diagrams of these processes as induced by neutrino
NSI are shown in Fig.~\ref{Fig:treelevel} for the $Z'$ model at tree level.
We include the interference with SM production as well.
At the LHC one will not be able to identify the flavor of neutrinos in
the final states.
We introduce
$\epsilon^{u(d)} \equiv \sum_{\alpha,\beta}|\epsilon_{\alpha\beta}^{u(d),V}|^{2}$
summed over all neutrino flavors.
It is understood that in case of $Z'$ model above couplings are constructed
out from the couplings with $Z'$ as in Eq.~(\ref{conv}).
The NSI contributions to cross sections at LHC thus are directly sensitive
to $\epsilon^{u(d)}$ with which we set the constraint
\footnote{In the actual calculation we assume only $\epsilon_{ee}^{u(d),V}$ are
non-zero and derive the constraint. However, since the
interference effects between NSI and SM interactions are small, one can
translate the same constraint to $\epsilon^{u(d)}$ as a good approximation.}.
The LHC measurements on NSI are complementary to those from neutrino oscillations
in the sense that they probe absolute values of the couplings rather than
differences of couplings of different flavors.
We present constraints on NSI in both frameworks of EFT and simplified
$Z'$ model.
It is understood that the former one equals the later constraint
with sufficiently large $M_{Z'}$.

For MC simulation of the NSI signals we use a model file
generated with FeynRule~\cite{Alloul:2013bka} similar to that used for dark matters with spin-1
mediator~\cite{Neubert:2015fka}.
We generate signal samples with
MG5\_aMC@NLO~\cite{MadGraph} followed by parton showering and
hadronizations with PYTHIA8~\cite{Sjostrand:2014zea},
and analyse the events with MadAnalysis5~\cite{Conte:2012fm}.
We use CTEQ6M~\cite{Pumplin:2002vw} PDFs in the simulation and use the default renormalization
and factorization scale choices in MadGraph5, which is the sum of transverse energy of
all final states divided by two.
In this section we report results using leading-order calculations matched
with parton showering and hadronization.
We will discuss the impact of next-to-leading order (NLO) QCD corrections and theoretical
uncertainties due to scale variation and choice of parton distribution functions
later.

\subsection{Data selection}

We summarize the LHC data sets used in our study.
They include recent measurements of mono-jet\cite{Aaboud:2017phn,Sirunyan:2017jix},
mono-V\cite{Sirunyan:2017jix,Aaboud:2018xdl}, mono-photon\cite{Aaboud:2017dor,Sirunyan:2018dsf},
and mono-lepton\cite{Sirunyan:2018mpc,Aad:2019wvl} production from both ATLAS
and CMS collaborations at LHC 13 TeV. 
The experimental analyses unfold the raw data to particle level with minimal
selection cuts.
The final measurements are presented in a model-independent form of upper limit
on total cross section in different fiducial regions.
That ensures a direct comparison to various new physics
beyond the standard model which generates large missing transverse momentums.
We reproduce the major selection cuts used in all analyses
as below for completeness.
In the following jets are clustered with anti-$k_T$ jet algorithm~\cite{0802.1189}
and a distance parameter of $D=0.4$ unless specified.
We start with measurements on hadronic final states recoiling against large missing energies.
The mono-jet production has the largest rate among all processes considered.
In the ATLAS analysis it requires a lower threshold on the missing transverse
momentum of $p_{T}^{miss}>$ 250 GeV.
For the visible objects it requires a leading jet with $p_{T}>$ 250 GeV
and $|\eta| < 2.4$, and a maximum of four jets with $p_{T} > 30$ GeV and
$|\eta| < 2.8$.
Furthermore, the separation of each jet and the missing transverse momentum in azimuthal plane should satisfy $\Delta\phi(j,\vec{p}_{T}^{miss})>$ 0.4. 
The CMS analysis imposes the same lower threshold of $p_{T}^{miss}>$ 250 GeV,
and requires a leading jet with $p_{T}>$ 100 GeV and $|\eta| < 2.4$.
The separation in azimuthal plane are $\Delta\phi(j,\vec{p}_{T}^{miss})>$ 0.5 for
each of the first four leading jets with $p_T>$ 30 GeV.
Unlike the ATLAS case no jet veto are applied in the CMS analysis.
For the production of large missing energies with a single $W/Z$ boson, and
subsequent hadronic decays, both ATLAS and CMS collaborations use sophisticated
technique of jet substructures to isolate the hadronic decaying $W/Z$ bosons
from backgrounds of QCD jets production.
However, efficiencies of those selections are derived for specific
models, and can be applied to deduce limits on cross sections of production
of $W/Z$ boson without decays.
In this sense the minimum requirements are a lower threshold of 250 GeV
for both the missing transverse momentum and the transverse momentum of
the $W/Z$ boson.
The ATLAS analysis presents results for final state with $W$ and $Z$
boson separately while CMS analysis only shows results with $W$ and
$Z$ boson combined.
In case of production of large missing energies with a $W$ boson, and
subsequent leptonic decays, that leads to the mono-lepton signatures.
Indeed such final states are indistinguishable from those induced by
production of a heavy $W'$ boson followed with leptonic decays.
The principal variable used in both ATLAS and CMS analyses concerning
mono-lepton signature is the transverse mass of the charged lepton
and the missing transverse momentum, $m_{T}$.
The ATLAS analysis requires electron (muon) candidates to have $|\eta|<1.37$
or $1.52<|\eta|<2.47$ ($|\eta|<2.5$) and $p_{T}>$ 65(55) GeV.
The lower threshold on missing transverse momentum $p_{T}^{miss}$ and
the transverse mass $m_{T}$ are 65 GeV and 130 GeV respectively for
electron final state, and 55 GeV and 110 GeV for muon.
The CMS analysis requires electron (muon) candidates to have $|\eta|<1.44$
or $1.56<|\eta|<2.47$ ($|\eta|<2.4$) and $p_{T}>$ 130(53) GeV.
A lower limit of 150 GeV on missing transverse momentum is imposed.
In the ATLAS mono-photon analysis it requires a leading photon with
$|\eta|<1.37$ or $1.52<|\eta|<2.37$ and $p_{T}^{\gamma}>$ 150 GeV,
and $\Delta \phi (\gamma , p_{T}^{miss})>$ 0.4.
The CMS analysis requires a leading photon with $p_{T}^{\gamma}>$ 175 GeV
and $|\eta|<1.44$.
In addition the missing transverse momentum should satisfy $p_{T}^{miss}>$ 170 
GeV and $p_{T}^{\gamma}$/$p_{T}^{miss}<1.4$.
In all above analyses the measured cross sections are binned in the
principal variables, which are $p_{T}^{miss}$, $p_{T}^{\gamma}$, and $m_{T}$
for mono-jet and mono-$V$, mono-photon, and mono-lepton, respectively.
Each bin in the principal variable is also called an exclusive region.
Besides, ATLAS and CMS also measure the cumulated cross sections
from a lower threshold of the principal variable to almost the largest
value allowed.
Each of those selected kinematic range is called an inclusive region.
We use the cross sections measured in inclusive regions to constrain
the non-standard interactions in our analysis by default, and compare
results to those obtained from exclusive regions if the latter is
available.
In Table.~\ref{tab:data} we summarize further information on the
LHC data sets used in our analysis.
That includes the total luminosity corresponds to each measurement,
the largest sensible values of the principal variable probed
in each measurement, and a ranking on different measurements
according to the constraint derived.
The CMS mono-jet measurement sets the strongest constraint on the 
non-standard interactions, followed by the CMS mono-$W/Z$ measurement,
ATLAS mono-jet and mono-$Z$ measurements.
\begin{table}[hpb]
  \centering
  \begin{tabular}{c|cccc}
  \hline
  Label & Variable & Range & Lum. (fb$^{-1}$) & Ranking\\
  \hline
      $CMS_{J}$ \cite{Sirunyan:2017jix}& $p_{T}^{miss}$ & 1250 GeV & 35.9 & 1 \\
      $CMS_{W+Z}$ \cite{Sirunyan:2017jix}& $p_{T}^{miss}$ & 750 GeV & 35.9 & 2 \\
      $ATLAS_{J}$ \cite{Aaboud:2017phn}& $p_{T}^{miss}$ & 1000 GeV & 36.1 & 3 \\
      $ATLAS_{Z}$ \cite{Aaboud:2018xdl}& $p_{T}^{miss}$ & 1500 GeV & 36.1 & 4 \\
      $CMS_{\gamma}$ \cite{Sirunyan:2018dsf}& $p_{T}^{\gamma}$ & 1000 GeV & 35.9 & 5 \\
      $ATLAS_{\gamma}$ \cite{Aaboud:2017dor}& $p_{T}^{\gamma}$ & 1000 GeV & 36.1 & 6 \\
      $ATLAS_{W}$ \cite{Aaboud:2018xdl}& $p_{T}^{miss}$ & 1500 GeV & 36.1 & 7 \\
      $ATLAS_{e}$ \cite{Aad:2019wvl}& $m_{T}^{min}$  & 5127 GeV & 139 & 8 \\
      $ATLAS_{\mu}$ \cite{Aad:2019wvl}& $m_{T}^{min}$  & 5127 GeV & 139 & 9 \\
      $CMS_{e}$ \cite{Sirunyan:2018mpc}& $m_{T}^{min}$ & 5127 GeV & 35.9 & 10 \\
      $CMS_{\mu}$ \cite{Sirunyan:2018mpc}& $m_{T}^{min}$ & 5127 GeV & 35.9 & 11 \\
 \hline
  \end{tabular}
  \caption{
	 Summary of various information on data sets used in this study, 
   including the principal variable used in each data set, its highest
   value probed, the total luminosity, and a ranking of different data sets.
	\label{tab:data}}
\end{table}
We explain briefly the statistical procedure used to derive exclusion limit on the
non-standard interactions.
We use the CLs\cite{AL2002Presentation} method together with the 
log-likelihood $\chi^2$ as a function of the model parameters and
the signal strength $\mu$,
\begin{equation}
\chi^{2}\left(\mu, \epsilon, M_{Z^{\prime}}\right)=\sum_{i=1}^{n}
\frac{\left(N_{\text {obs},i}-N_{bg,i}-\mu \sigma_{i}(\epsilon ,
	M_{Z^{\prime}}) \mathcal{L}\right)^{2}}{N_{\text {obs},i}+
\delta_{sys,i}^{2}}=\chi^{2}_{0}+A\mu+B\mu^{2},
\label{eq:chi}
\end{equation}
for each data set and with $i$ runs from all regions considered.
For each region, $N_{obs,i}$ and $N_{bg,i}$ are the total number of events
observed and predicted by the SM, $\delta_{sys,i}$ is the
total systematic error, and $\sigma_{i}(\epsilon , M_{Z^{\prime}})$
represents the cross section predicted by the model of non-standard
interactions.
$\mathcal {L}$ is the integrated luminosity.
The $\chi^2$ is a quadratic function of $\mu$ with coefficients
$A$, $B$ and $\chi^{2}_{0}$ depending on model parameters $M_{Z^{\prime}}$
and $\epsilon$. 
CLs upper limit on the NSI strength $\epsilon$ for fixed
$M_{Z^{\prime}}$ at a confidence
level 1-$\alpha^{\prime}$ is determined by
\begin{equation}
\hat{\mu}+\Delta_{\mu} \Phi^{-1}\left(1-\alpha^{\prime} \Phi(\hat{\mu}
/ \Delta_{\mu})\right)=1,
\label{eq:llhood}
\end{equation}
with $\hat{\mu}= -{A}/{2B}$, $\Delta_{\mu}=1/\sqrt{B}$.
$\Phi$ is the cumulative distribution function of normal distribution.
In case of using exclusive region/bin we can include all regions of
the data set into $\chi^2$ to derive the limit on $\epsilon$ if bin-bin
experimental correlations are known.
For using inclusive region, we can only include one of them at a time
since different inclusive regions are statistically correlated.
For a single inclusive region/bin, the CLs limit on the cross section
induced by non-standard interactions can be simplified as
\begin{equation}
	\sigma_{\mathrm{up}}=\hat{\sigma}+\Delta_{\sigma} \Phi^{-1}\left(1-\alpha^{\prime}
	\Phi(\hat{\sigma} / \Delta_{\sigma})\right),
\label{eq:singlebin}
\end{equation}
with the maximum likelihood estimator and the uncertainty of $\sigma$
given by
\begin{equation}
\hat{\sigma}=(N_{obs}-N_{bg})/\mathcal{L},\qquad
\Delta_{\sigma}=\sqrt{N_{obs}+\delta_{sys}^{2}}/\mathcal{L}.
\end{equation}
In our analysis for each data set we scan over all the inclusive
regions for the exclusive limit on $\epsilon$ and take the strongest
one among them.
We have verified explicitly with the CMS mono-jet measurement that the exclusion
limit as derived from a scan on the inclusive regions is similar to that
obtained using a $\chi^2$ with all exclusive regions.

\subsection{Constraints for effective operator}

\begin{figure}[htbp]
	\centering
	\includegraphics[width=.7\textwidth,clip]{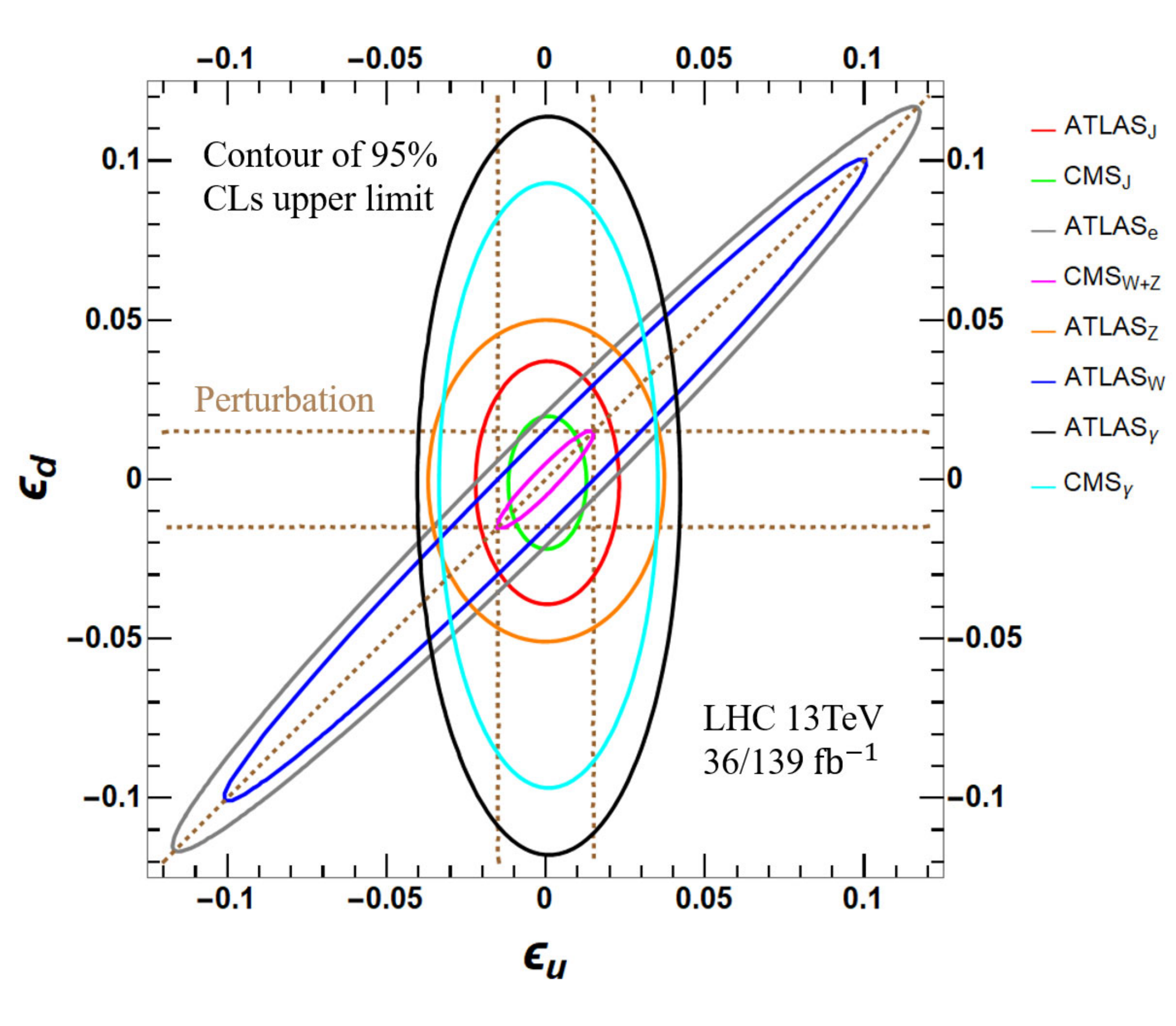}
	\caption{%
		Contour of 95\% CLs upper limit on the plane of NSIs
		$\epsilon_u$ and $\epsilon_d$ in the framework of effective
		operators with various measurements at
		the LHC. 
		The vertical and horizontal dashed lines indicate bounds
		from perturbative conditions.
		}
	\label{Fig:100TeV}
\end{figure}

We first present results for case of using effective operator.
In Fig.~\ref{Fig:100TeV} we show 
the contour of 95\% CLs upper limit on the plane of the NSI $\epsilon_u$
and $\epsilon_d$,
from all LHC data sets
discussed earlier.
We only include one representative result for mono-lepton from CMS for simplicity.
The effective parton-parton center-of-mass energy is approximately
5 TeV for 13 TeV run of LHC.
It indicates the new physics scale $\Lambda$ in the effective operator
approach should be larger than that to ensure its validity.
Meanwhile, the Wilson coefficient $c$ in Eq.(\ref{EFToperator}) can not exceed
a perturbative bound of $4\pi$ assuming it is induced by tree-level amplitude in
a weakly coupled theory.
That sets a boundary value of about $0.015$ for the NSIs as shown by
the dashed horizontal and vertical lines in Fig.~\ref{Fig:100TeV}.
Outside the bounded region the effective operator approach is not
valid at the LHC if one does not apply any cut on the center-of-mass
energy of the scattering.
Furthermore, for NSIs with left-handed quarks, to maintain the
gauge invariance of SM $SU(2)_L$ symmetry one should set
$\epsilon_u=\epsilon_d$.
Otherwise it may lead to apparently too strong constraints on
the NSIs in the direction of $\epsilon_u=-\epsilon_d$ due to
violation of gauge invariance~\cite{Bell:2015rdw}, as can be seen for mono-$W$ and
mono-lepton production in Fig.~\ref{Fig:100TeV}.
In the following we will focus on constraints along diagonal
direction $\epsilon_u=\epsilon_d=\epsilon$.

The CMS mono-jet measurement sets the strongest constraint of
$\epsilon\lesssim 0.011$, followed by CMS mono-$W/Z$ with constraints
of $\epsilon\lesssim 0.015$, both within the perturbative region,
and ATLAS mono-jet with constraints of $\epsilon\lesssim 0.020$.
Measurements on mono-photon and mono-lepton production lead to much
weaker constraints.
The constraints are almost symmetric in the positive and negative
directions due to the relatively smallness of interference effects with SM for
the NSI strength probed.
The asymmetry can be measured by the difference of $\epsilon^{+}$ and $\epsilon^{-}$, which are bounds in the positive and negative directions respectively. 
For the case of CMS mono-jet production, we have $(|\epsilon^{+}|-|\epsilon^{-}|)/|\epsilon^{+}| \sim$ 2\%.
It is interesting that CMS measurements in general impose stronger
constraints than ATLAS for the same final states due to
the smaller systematic uncertainties of CMS.

\subsection{Constraints for simplified $Z'$ model}
\begin{figure}[htbp]
	\centering
	\includegraphics[width=.8\textwidth,clip]{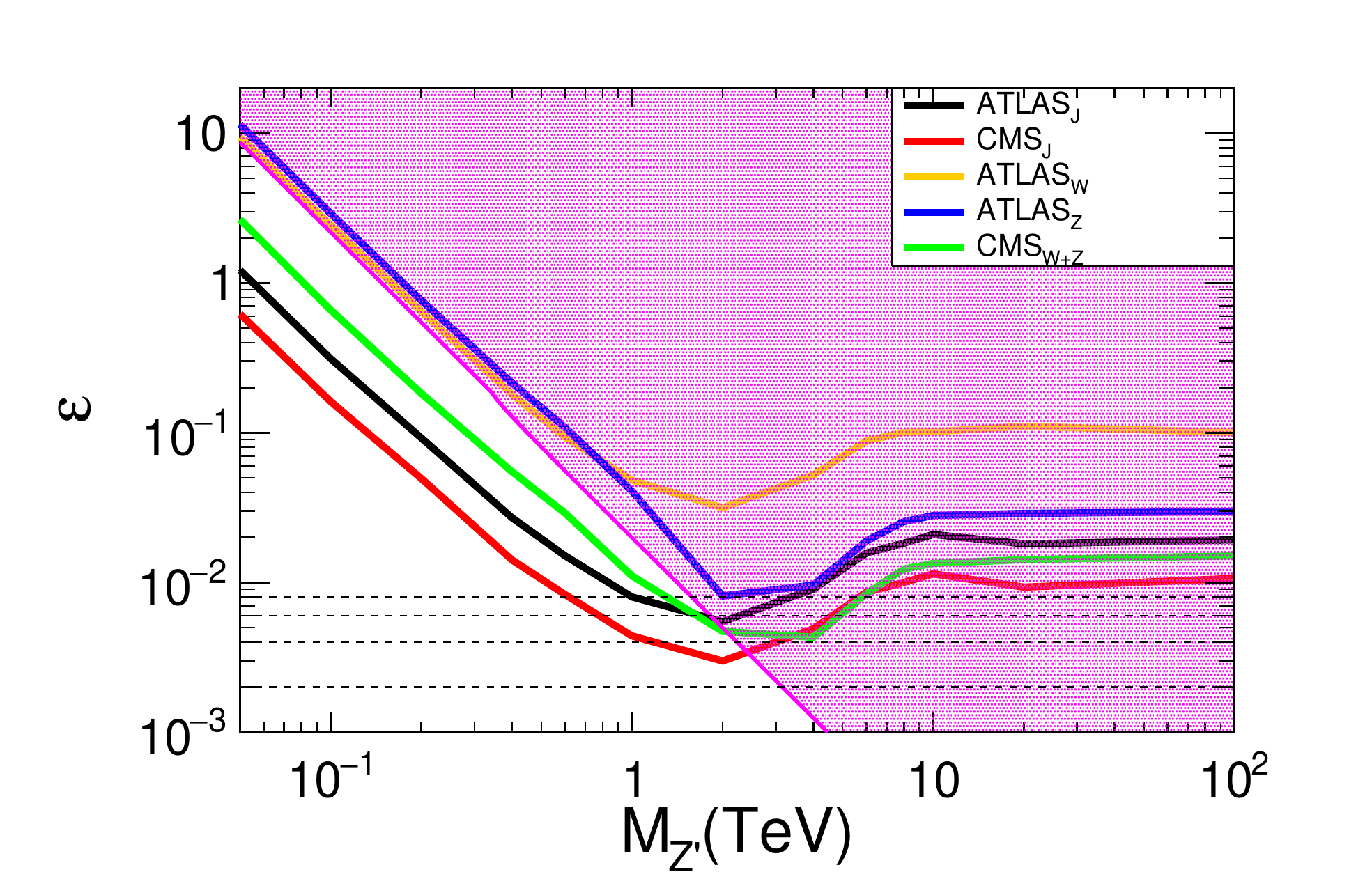}
	\caption{
		95\% CLs upper limit on NSIs in a simplified $Z'$ model
		as a function of the mass of $Z'$ with various measurements at
		the LHC. 
	We assume $ \epsilon_{u} = \epsilon_{d}=\epsilon$
	and $\Gamma_{Z^{\prime}}/M_{Z^{\prime}}=0.1$.}
	\label{Fig:M_eps}
\end{figure}

\begin{figure}[tbp]
	\centering
	\includegraphics[width=.49\textwidth,clip]{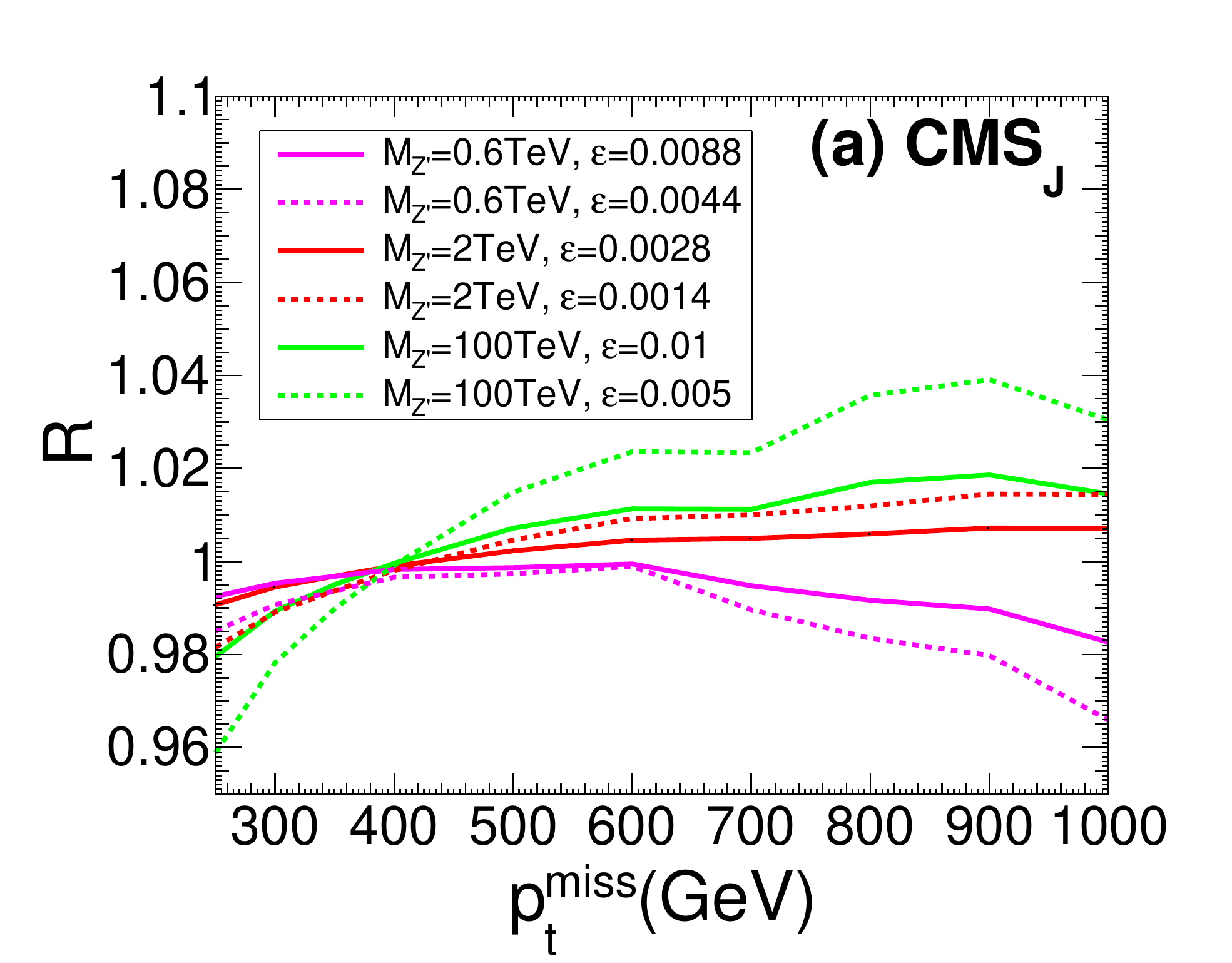}
	\hfill		
	\includegraphics[width=.49\textwidth,clip]{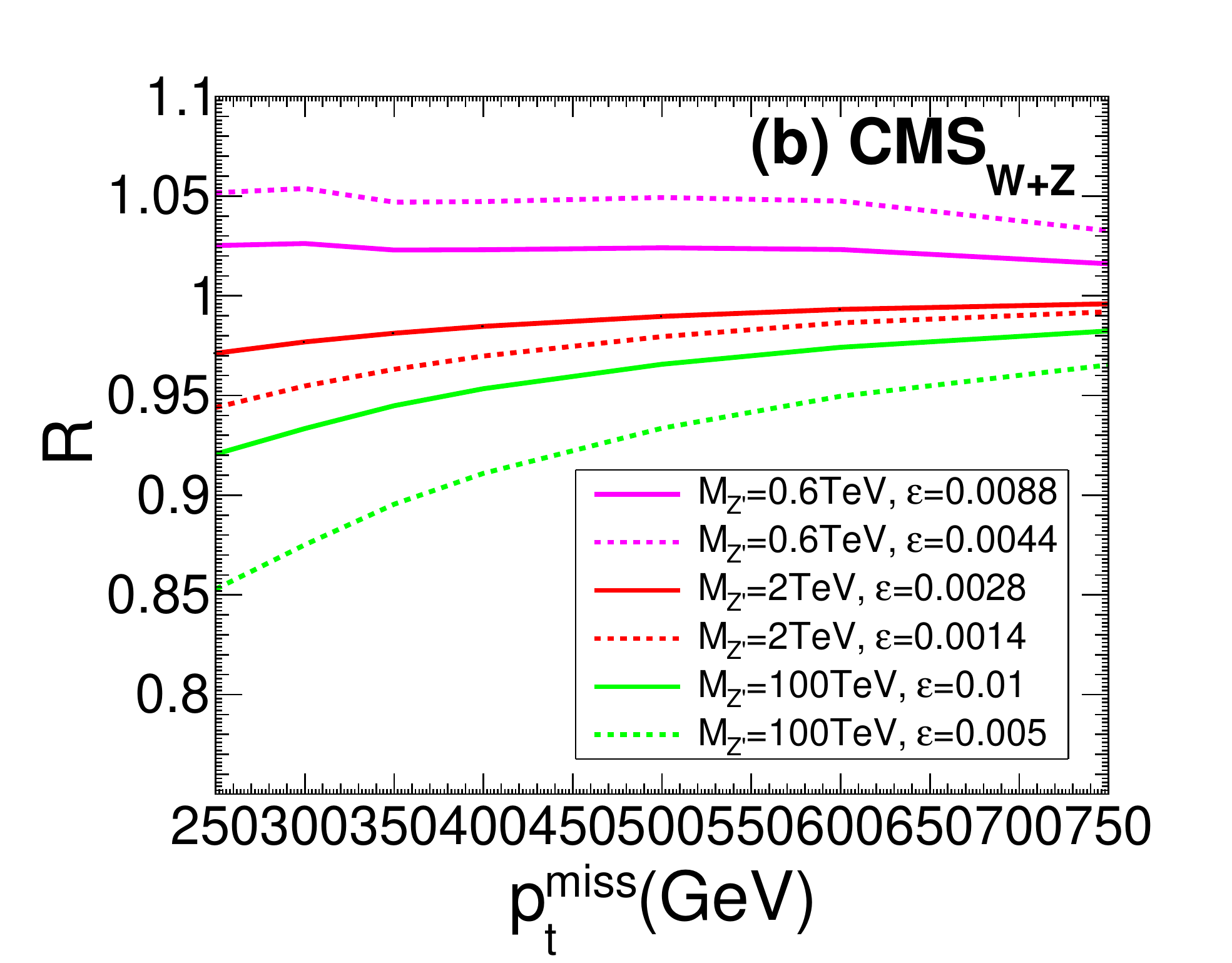}
	\hfill
  \caption{
	  Ratio of predicted cross-sections without and with interference
	  contributions, for production of CMS mono-jet and mono-$W/Z$ on
	  the left and right respectively.
	  Results are shown for a few benchmark points of $M_{Z'}$ and $\epsilon$
	  according to current and future LHC sensitivity.}
  \label{Fig:inter}
\end{figure}

We turn to the constraints for NSIs from simplified $Z'$ model.
The production cross sections at the LHC depend on couplings of
the $Z'$ boson to quarks, neutrinos, as well as on mass and width
of the boson, $M_{Z'}$ and $\Gamma_{Z'}$.
Similar as before we would like to translate the constraints to
the conventional NSI parameter $\epsilon_u=\epsilon_d=\epsilon$ defined in Eq.~(\ref{conv}).
For a fixed value of $\epsilon$, one can still vary $M_{Z'}$
and $\Gamma_{Z'}$ for changes of cross sections at the LHC.
We fix $\Gamma_{Z'}/M_{Z'}=0.1$ and study the constraints on $\epsilon$
as a function of the mass of $Z'$ boson for simplicity.
In Fig.~\ref{Fig:M_eps} we summarize the constraints imposed by mono-jet
and mono-$V$ measurements from both ATLAS and CMS.
We show the 95\% CLs upper limits on the positive axis of $\epsilon$.
The limits on negative side of the axis are quite similar since the
interference effects are small in general.
For example, in CMS mono-jet production with $M_{Z'}=$ 2 TeV, the
value of $(|\epsilon^{+}|-|\epsilon^{-}|)/|\epsilon^{+}|$ is about 0.5\%.
The impact of interference between NSIs and SM can be further visualized in
Fig.~\ref{Fig:inter}, where we show ratios between predicted cross-sections
without and with including interference contributions.
We choose a few benchmark points of $M_{Z'}$ and $\epsilon$ that are close
to the current and future sensitivity of LHC.
We can see for the large transverse momentum region, which has the strongest
sensitivity to NSIs, the interference terms contribute less than 10\% to the
cross sections.
For a certain choice of $M_{Z'}$ the NSI $\epsilon$ can not be arbitrarily 
large otherwise the partial widths of $Z'$ decaying into neutrinos and
quarks can easily saturate the assumed total width.
That leads to a theoretical upper bound on the NSI as~\cite{Babu:2020nna}
\begin{equation}
|\epsilon|\leq \dfrac{\sqrt{3}\pi}{\sqrt{N}G_{F}M_{Z^{\prime}}^{2}}
\dfrac{\Gamma_{Z^{\prime}}}{M_{Z^{\prime}}},
\end{equation}
where $N$ is the number of massless quarks plus possible contributions
from heavy quarks with mass below $M_{Z'}/2$.
The parameter space in Fig.~\ref{Fig:M_eps} with colors are thus excluded.

The CMS measurement on mono-jet production again gives the strongest
constraints, with an upper limit ranging from about 0.6 for a $Z'$
mass of 50 GeV to 0.0028 for a $Z'$ mass of 2 TeV.
The constraints become weaker when $M_{Z'}$ goes beyond 2 TeV since
then the $Z'$ boson can hardly be produced directly. 
The ATLAS mono-jet measurement sets a limit of about two times
larger than CMS.
Our results on constraints from ATLAS mono-jet production agree well with that
shown in Ref.~\cite{Babu:2020nna}.
The constraints from CMS mono-$V$ measurement are weaker than
those from ATLAS mono-jet except for very large $M_{Z'}$. 
The constraints from  ATLAS mono-$Z/W$ measurements are weaker
than the theoretical bounds.
Interestingly, results for the $Z'$ model approach smoothly to those of
effective operators with increasing $Z'$ mass, as demonstrated in
Fig.~\ref{Fig:M_eps}.
In all cases the limits are strongest for a $Z'$ mass of about 2 TeV, and
increase afterwards, and finally stabilize for $M_{Z'} \gtrsim$ 6 TeV. 
Results shown in Fig.~\ref{Fig:M_eps} can also be translated into constraints
for different choices of $\Gamma_{Z'}/M_{Z'}$ easily.
For instance, if instead assuming $\Gamma_{Z'}/M_{Z'}=$ 5\%, constraints
from all data sets will scale down by $1/\sqrt 2$ since the cross sections
are approximately proportional to $\epsilon^2/\Gamma_{Z'}$ for not too heavy
$Z'$.
Meanwhile, the theoretical bounds will scale down by a factor of 2 and
are more closer to the experimental constraints.

\section{Theoretical uncertainties}\label{sec:mass}

In this section we extend our results by using theoretical predictions
calculated at next-to-leading order in QCD.
The NLO QCD corrections can be potentially large in tail region of
various distributions, that have the strongest sensitivity to NSIs.
We further study impact of theoretical uncertainties on the
constraints to NSIs, including the scale variations and uncertainties
due to parton distribution functions.

\subsection{Next-to-leading order QCD corrections}

\begin{figure}[tbp]
	\centering
	\includegraphics[width=.49\textwidth,clip]{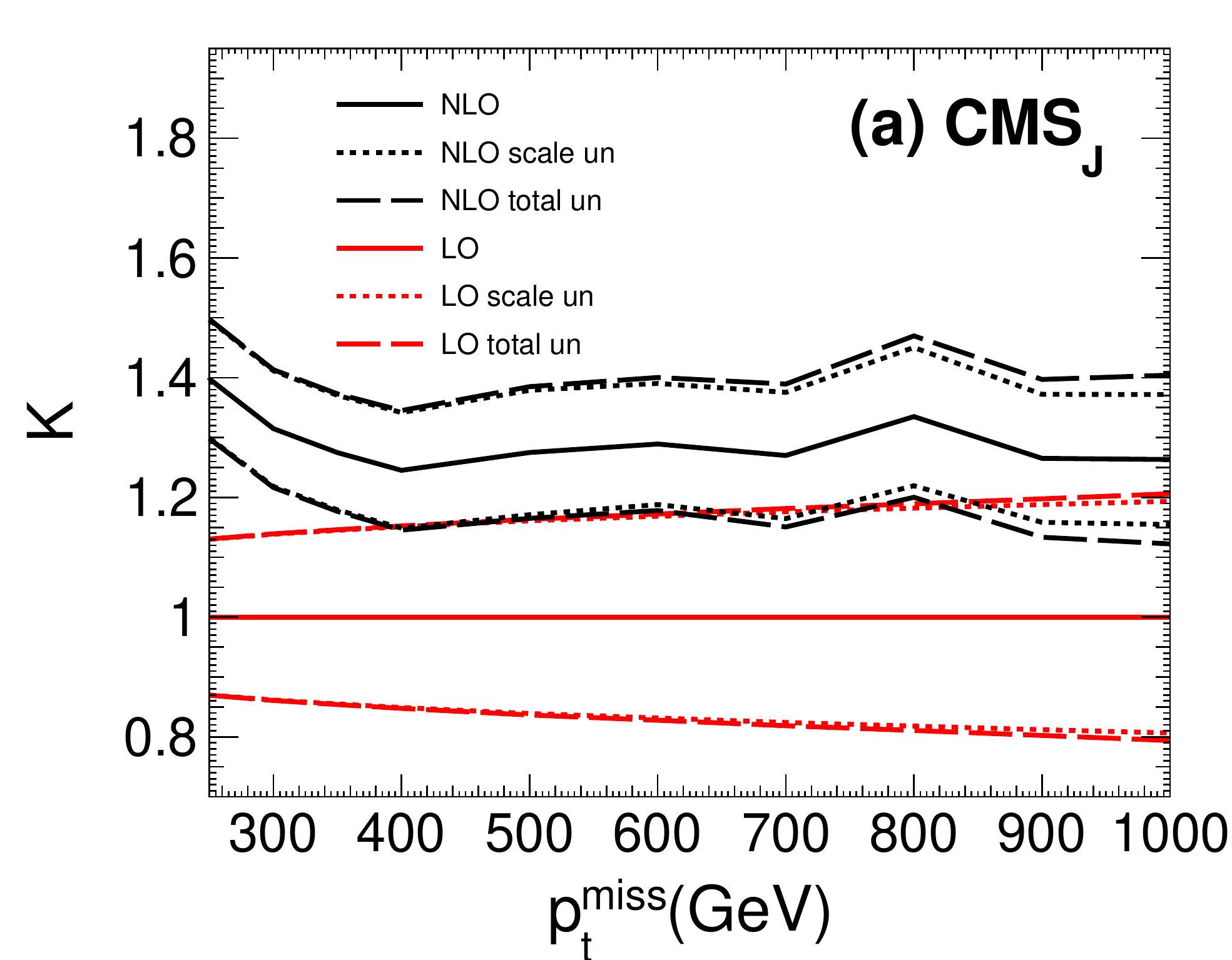}
	\hfill		
	\includegraphics[width=.49\textwidth,clip]{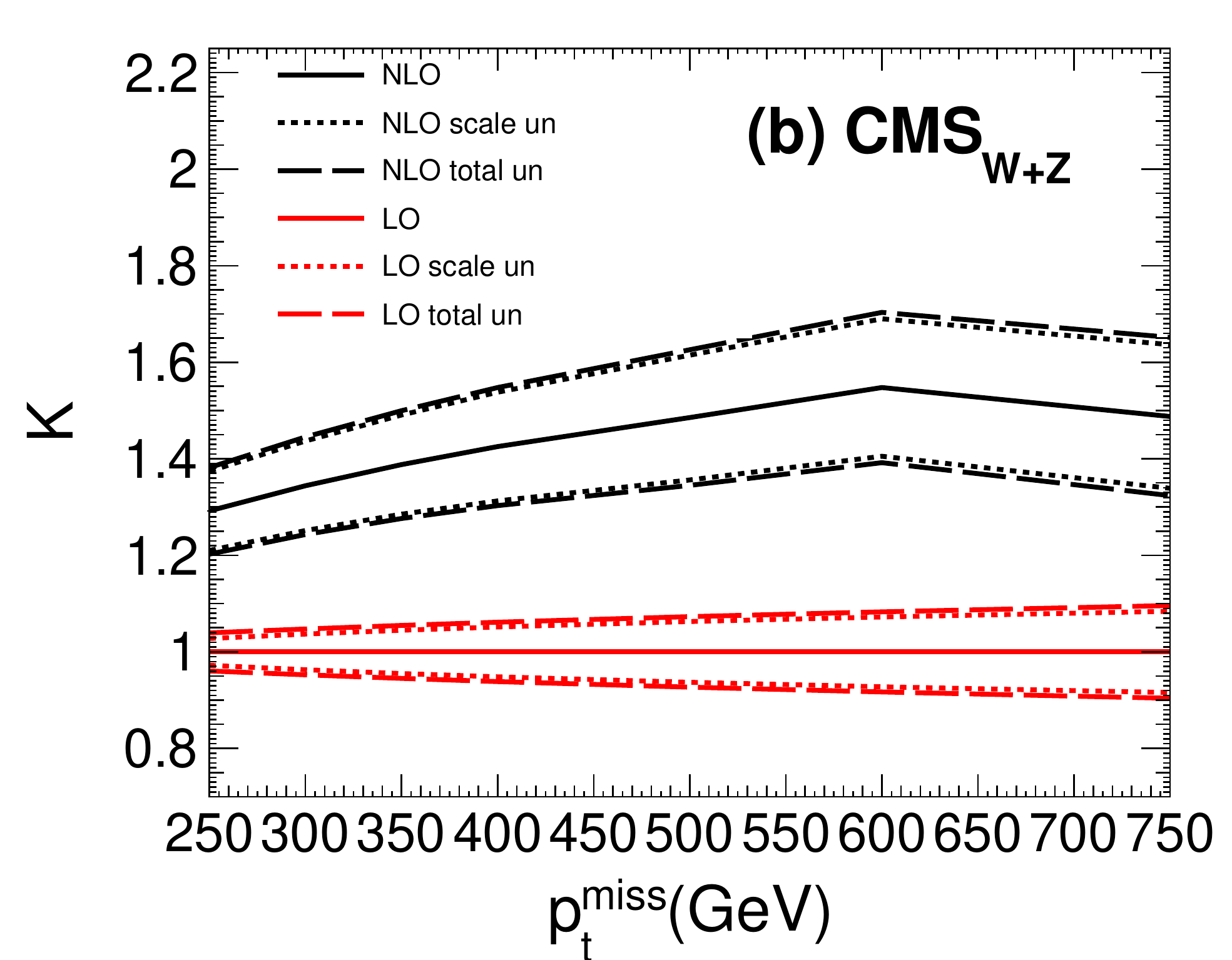}
	\hfill
  \caption{
	  K-factors defined as ratio of NLO cross sections to LO cross
	  sections for inclusive regions as a function of the lower threshold
	  for CMS mono-jet and mono-$W/Z$ on the left and right respectively,
	  for a $Z'$ mass of 100 GeV.
	  The dashed (dotted) lines show the total (scale) uncertainties
	  of the LO and NLO predictions. 
	  }
  \label{Fig:Kfactor1}
\end{figure}

\begin{figure}[tbp]
	\centering
	\includegraphics[width=.49\textwidth,clip]{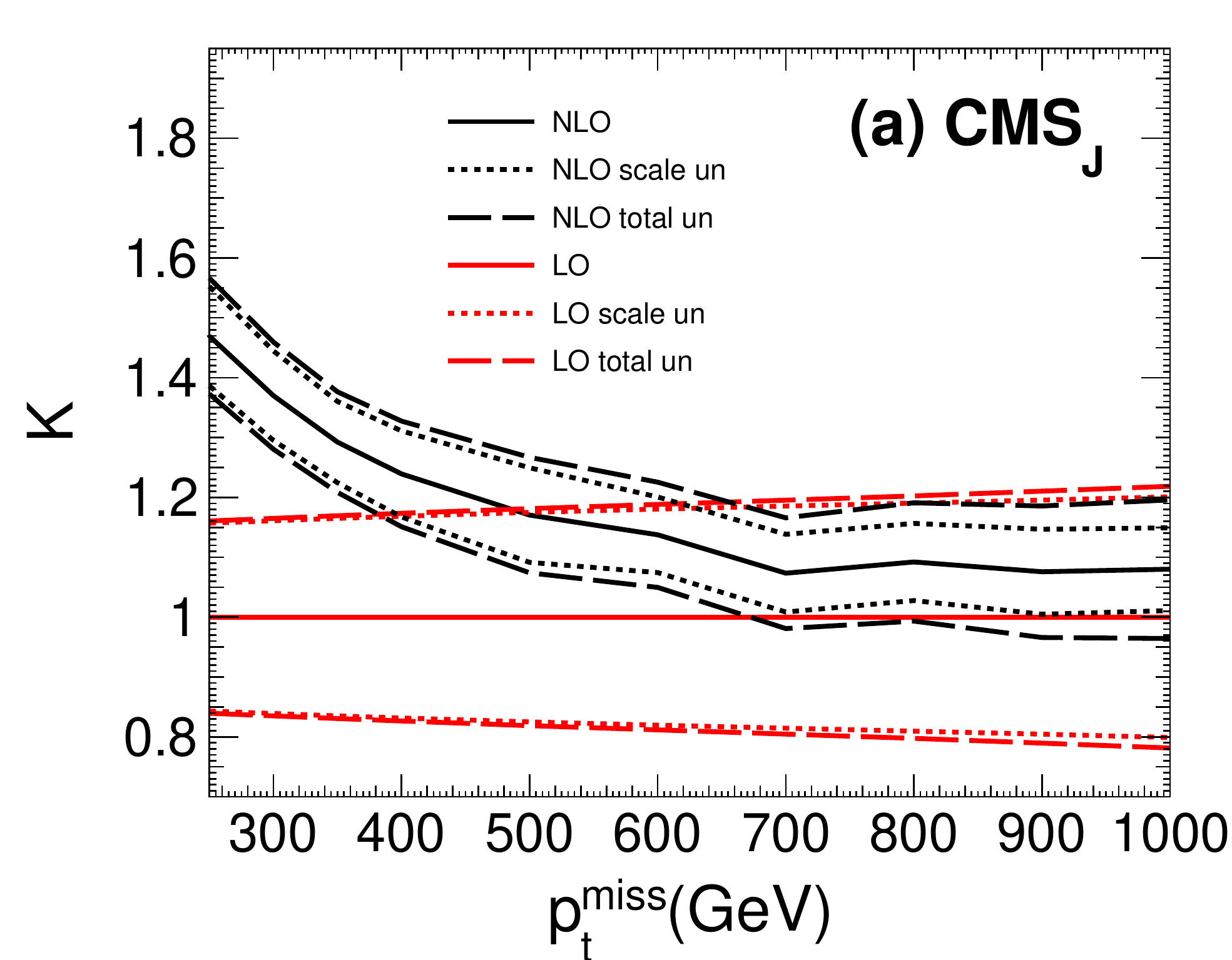}
	\hfill		
	\includegraphics[width=.49\textwidth,clip]{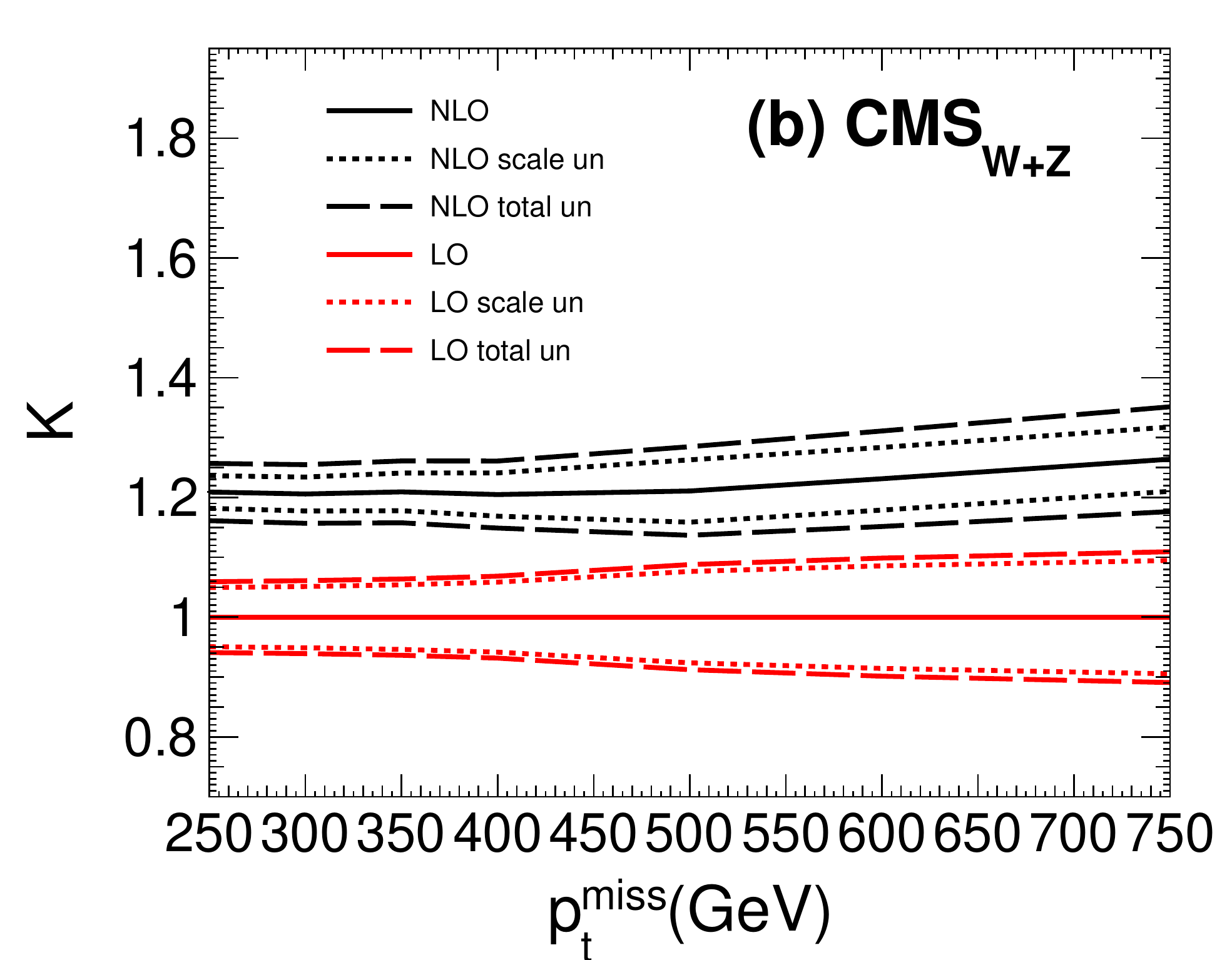}
	\hfill
	\caption{
	  K-factors defined as ratio of NLO cross sections to LO cross
	  sections for inclusive regions as a function of the lower threshold
	  for CMS mono-jet and mono-$W/Z$ on the left and right respectively,
	  for a $Z'$ mass of 1 TeV.
	  The dashed (dotted) lines show the total (scale) uncertainties
	  of the LO and NLO predictions. 
	  }
	\label{Fig:Kfactor2}
\end{figure}

The NLO calculations for various processes mentioned
earlier can be carried out straightforwardly by generating
the model file of NSIs at NLO in QCD with FeynRules~\cite{Alloul:2013bka}
followed by simulation with MG5\_aMC@NLO~\cite{MadGraph} and PYTHIA8~\cite{Sjostrand:2014zea}. 
The impact of corrections to various distributions can be
described by a K-factor defined as
\begin{equation}
	K(O_0) = \frac{\sigma_{NLO}(O>O_0)}{\sigma_{LO}(O>O_0)},
\end{equation}
calculated for different inclusive regions, where the numerator
and denominator are cumulated cross sections at NLO and LO respectively.
Our nominal predictions are calculated with the default choice
of QCD renormalization and factorization scales, and with
the central set of CTEQ6M NLO PDFs~\cite{Pumplin:2002vw}.
We vary the renormalization and factorization scales independently
by a factor of two and take the 9-scale envelope as the uncertainty
range.
The PDF uncertainties are calculated with Hessian error sets
provided in CTEQ6M PDFs at 68\% C.L.
The total theoretical uncertainties are quadratic sum of the
scale and PDF uncertainties in both plus and minus directions.
In Figs.~\ref{Fig:Kfactor1} and \ref{Fig:Kfactor2} we plot the K-factor
as a function of the lower threshold of principal observable for CMS mono-jet production
and CMS mono-$W/Z$ production, with the $Z'$ mass of 100 GeV
and 1 TeV respectively.
NLO corrections and theoretical uncertainties are quite similar for
the case of ATLAS mono-jet and mono-$V$ which we do not show for simplicity.
The solid line represents the K-factor and the dashed (dotted) lines
show the total (scale) uncertainty for LO and NLO predictions.
The QCD corrections start from 40 (45)\% at low $p_T^{miss}$ and decrease
to about 25 (10)\% for mono-jet production with $M_{Z'}=$100 GeV (1 TeV).
For mono-$W/Z$ production the QCD corrections are about 30 (20)\% at
low $p_T^{miss}$ and increase to about 50 (25)\% slowly with $M_{Z'}=$100 GeV (1 TeV).
The peculiar shape of K-factor in mono-jet plot with $M_{Z'}=$ 100 GeV
is partly due to the MC errors.
In all cases uncertainties due to scale variations are dominant over
PDF uncertainties.
We observe a reduction of scale uncertainties for NLO predictions
except for mono-$W/Z$ production with $M_{Z'}=$ 100 GeV where the
scale variations at LO underestimate the genuine perturbative
uncertainties.
The total uncertainties increase with $p_T^{miss}$ at both LO and NLO.
For NLO predictions the relative total uncertainties range between
7$\sim$11\% for mono-jet production with two choices of masses, and
between 4$\sim$10\% for mono-$W/Z$ production.

\subsection{Constraints at NLO}

\begin{figure}[tbp]
	\centering
	\includegraphics[width=.49\textwidth,clip]{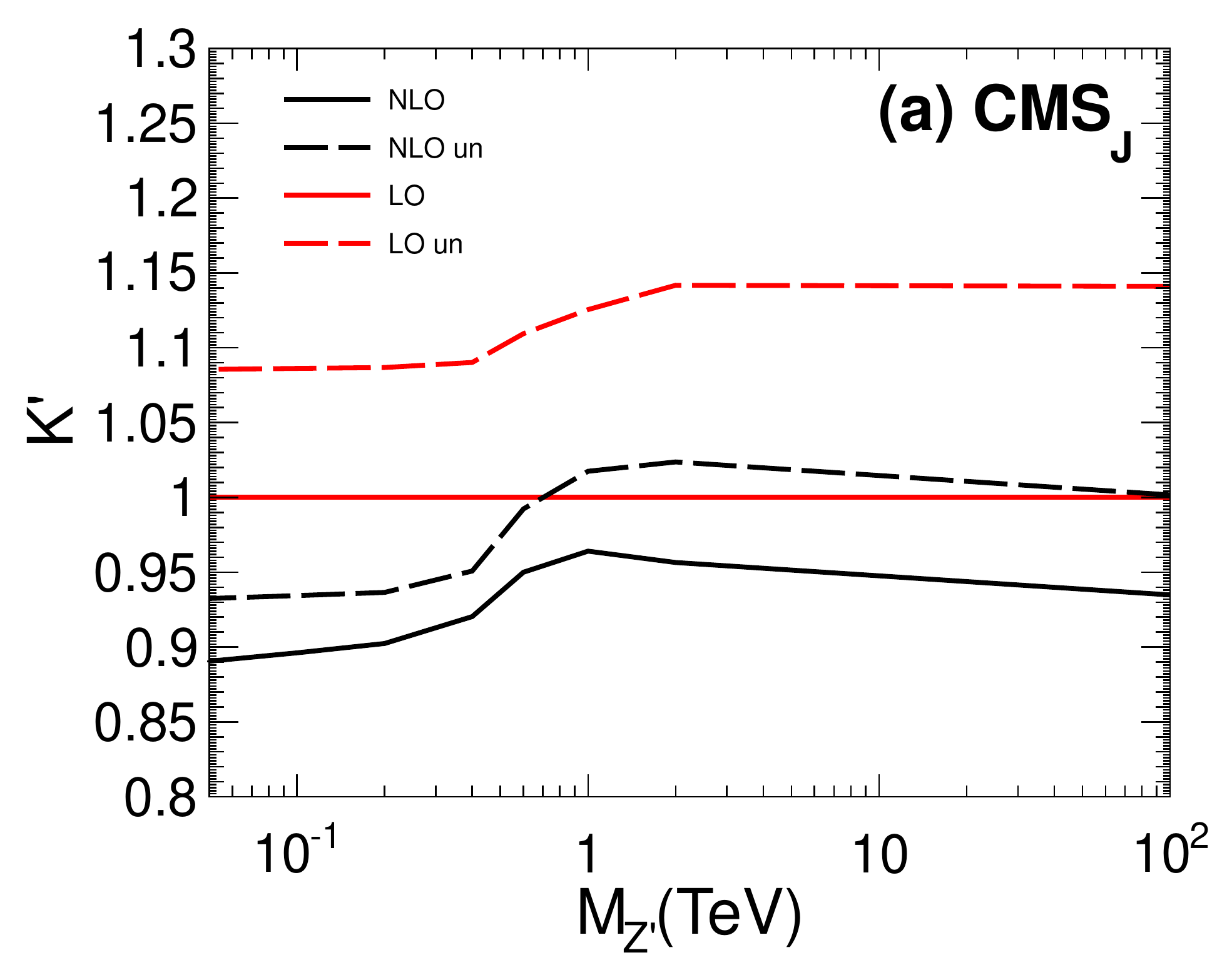}
	\hfill		
	\includegraphics[width=.49\textwidth,clip]{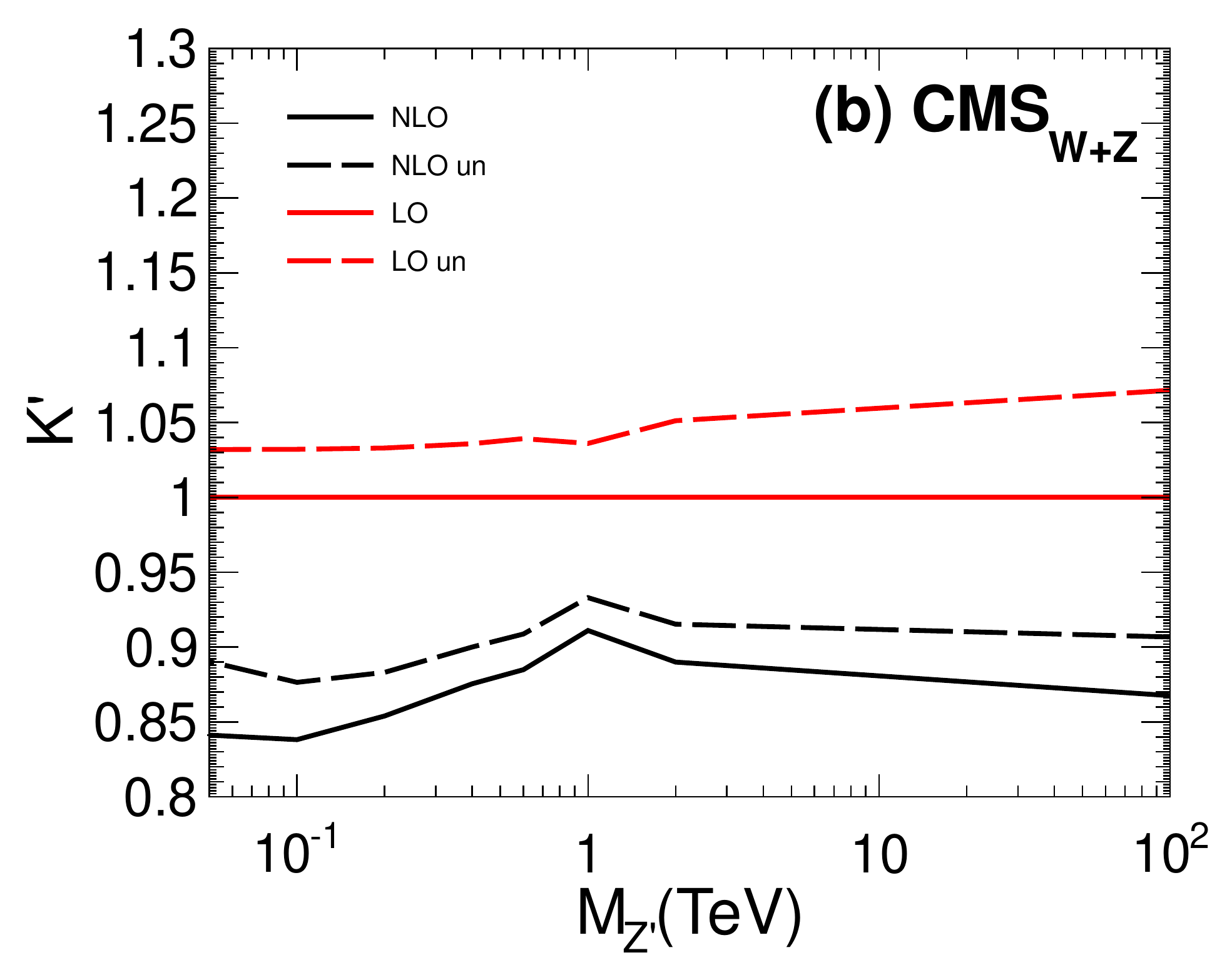}
	\hfill
  \caption{
	  Exclusion limits on NSIs using different theoretical predictions
	  normalized to those using leading order predictions without theoretical
	  uncertainties, as a function of the mass of $Z'$.
	  The left (right) plots corresponds to constraints from CMS mono-jet
	  (mono-$W/Z$) measurement.
	  }
  \label{Fig:Kprime}
\end{figure}

We are now ready to study impact of the NLO corrections and 
theoretical uncertainties on the derived limit of NSIs.
The results are presented in Fig.~\ref{Fig:Kprime} as a function
of the mass of $Z'$ for constraint with CMS mono-jet and
mono-$W/Z$ production respectively.
We derive four groups of 95\% CLs upper limit on $\epsilon$.
They include the two using our nominal LO and NLO predictions
on the cross sections.
In the other two we use the LO predictions scaled to the
lower side of the uncertainty band and similar for NLO predictions,
which corresponds to conservative constraints on NSIs than those
using nominal theory predictions.
We plot all four constraints normalized to the one with nominal
LO predictions as a function of $M_{Z'}$ in Fig.~\ref{Fig:Kprime}.
For the case of CMS mono-jet production, the theoretical uncertainties
weaken the limit by $10\sim 15$\% at LO across the full range of $M_{Z'}$.
The NLO corrections increase the cross sections and thus lead to
an improvement of $5\sim 10$\% on the constraints of NSIs.
The theoretical uncertainties have less impact at NLO than at LO
due to the stabilization of theory predictions at higher orders.
In combination with NLO corrections and theory uncertainties
the constraints on NSIs improve slightly as comparing to the
nominal LO ones that we show in Sect.~\ref{sec:aa}.
For CMS mono-$W/Z$ production, the theoretical uncertainties
change the limit by less than 10\% at LO and even smaller at NLO.
The constraints on NSIs are improved by 10\% in the full range
of $M_{Z'}$ when considering the NLO corrections together with
theoretical uncertainties.

\section{LHC combination and projections}\label{sec:thunc}

\begin{figure}[htbp]
	\centering
	\includegraphics[width=.8\textwidth,clip]{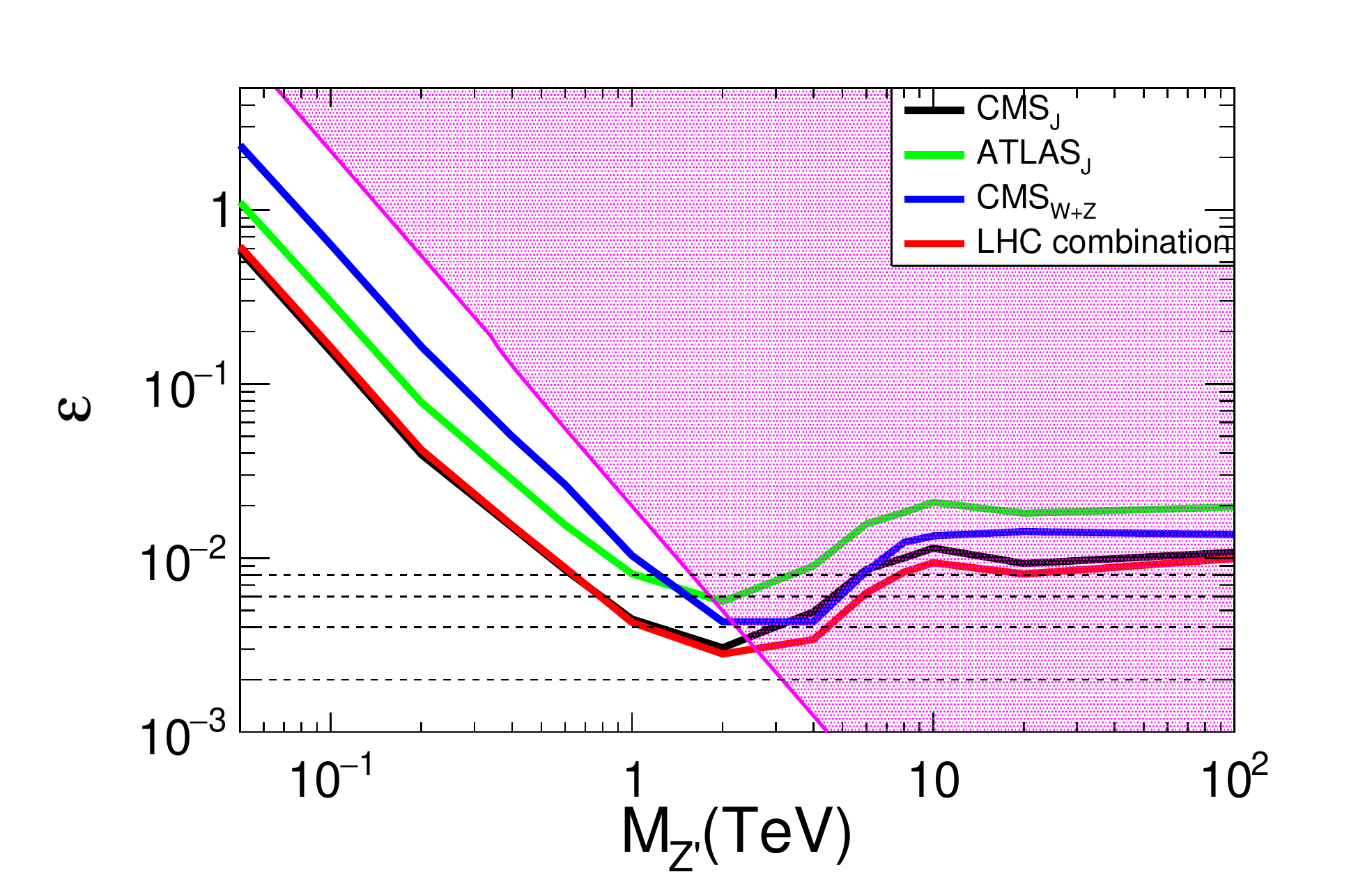}
	\caption{
		95\% CLs upper limit on NSIs in a simplified $Z'$ model
		as a function of the mass of $Z'$ with various measurements at
		the LHC and their combinations. 
	We assume $ \epsilon_{u} = \epsilon_{d}=\epsilon$
	and $\Gamma_{Z^{\prime}}/M_{Z^{\prime}}=0.1$, and use
	the NLO predictions with theoretical uncertainties.}
	\label{Fig:combine}
\end{figure}

We have shown that for individual measurement the strongest constraints
on NSIs arise from CMS mono-jet production in both the EFT framework and
the simplified $Z'$ model.
It is worth to study the improvement once we combine
constraints from several data sets, specifically the CMS mono-jet,
CMS mono-$W/Z$, and ATLAS mono-jet measurements.
For each value of the $Z'$ mass, we first identify the most sensitive
inclusive region for each of the three measurements.
We construct the total $\chi^2$ function in Eq.~(\ref{eq:chi}) as
a sum of the three individual $\chi^2$.
The 95\% CLs upper limit is then determined following the
prescription outlined earlier.
We neglect correlations between systematic errors of different
measurements which are not available.
The results are presented in Fig.~\ref{Fig:combine} using the
NLO predictions with theoretical uncertainties.
For $M_{Z'}$ below 1 TeV the combined limits are almost identical
to those from CMS mono-jet alone since the latter are stronger
by more than a factor of two than other data sets.
The constraints are improved by at least about 10\% for $M_{Z'}$ greater 
than 1 TeV.

\begin{figure}[tbp]
	\centering
	\includegraphics[width=.8\textwidth,clip]{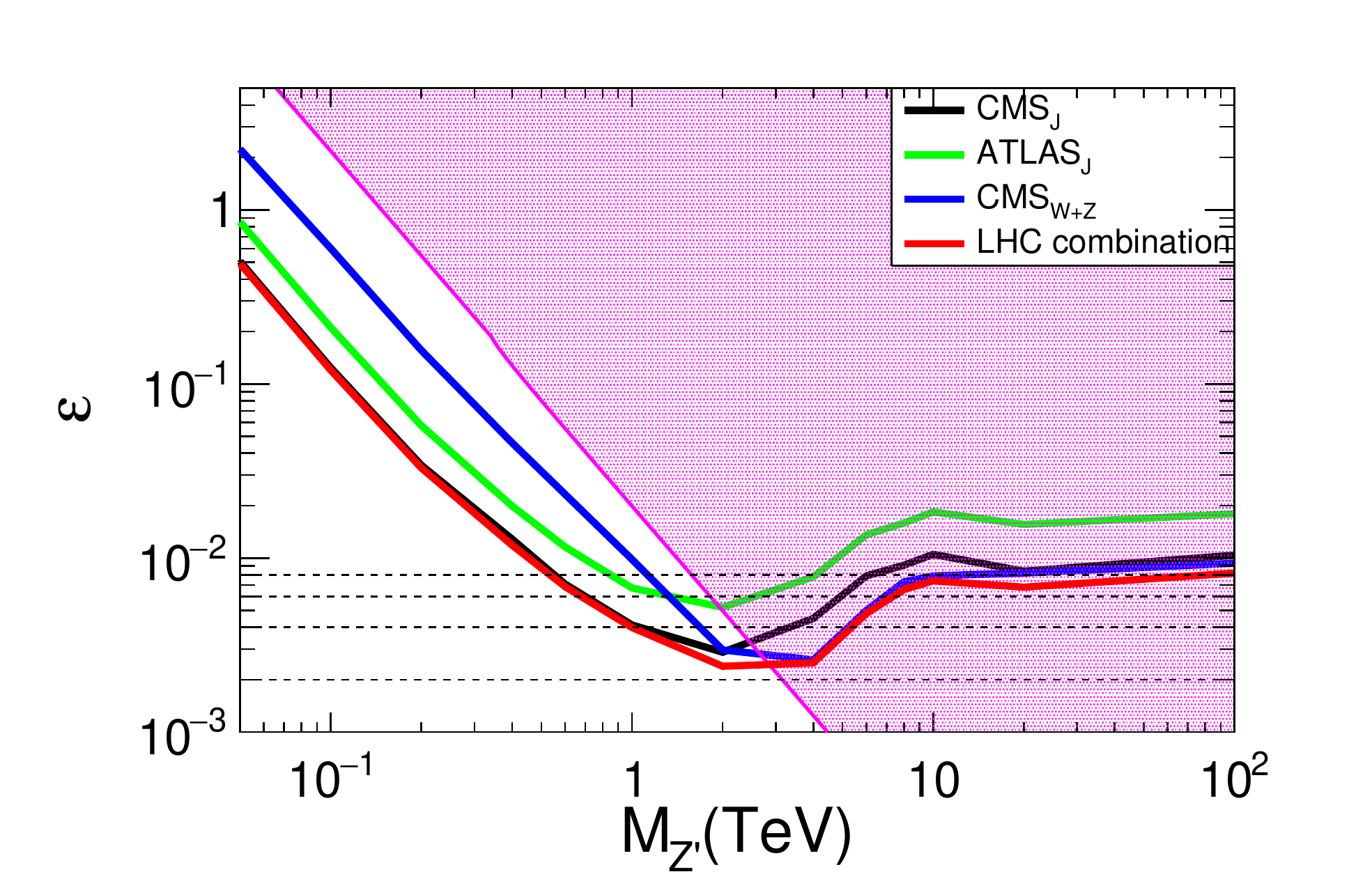}
	\hfill
  \caption{
		Projection of exclusion limit on NSIs for LHC with an integrated luminosity of
		300 fb$^{-1}$, in a simplified $Z'$ model
		as a function of the mass of $Z'$.
	We assume $ \epsilon_{u} = \epsilon_{d}=\epsilon$
	and $\Gamma_{Z^{\prime}}/M_{Z^{\prime}}=0.1$, and use
	the NLO predictions with theoretical uncertainties.}
  \label{Fig:HLLHC}
\end{figure}

The LHC is expected to accumulate a total integrated luminosity of
3000 fb$^{-1}$ for the high luminosity run.
The constraints on NSIs can benefit from the high statistics of various
measurements.
We calculate the projections for constraints on NSIs with mono-jet and mono-$W/Z$
measurements at LHC with higher luminosities.
We rescale the number of SM background events from current values
with luminosities and set the number of observed events to be the same as
the SM backgrounds.
We assume the relative size of systematic uncertainties remain the
same though one may expect certain improvements from both experimental
and theoretical sides.
In Fig.~\ref{Fig:HLLHC} we plot the expected upper limit on the
NSIs as a function of $M_{Z'}$ for an integrated luminosity of 300 fb$^{-1}$.
The increase of statistics improves the constraints as from CMS mono-$W/Z$
production while has less impact on mono-jet production since the
measurements are already dominated by systematic uncertainties.
It is interesting to find the constraints from CMS mono-$W/Z$ measurement
become as good as those from CMS mono-jet measurement for $M_{Z'}>$ 2 TeV.

We summarize the limits on $\epsilon$ in Table.~\ref{Tab:HLLHC} for several
choices of the mass of $Z'$ and different measurements.
It is understood that the case with $M_{Z'}=$ 100 TeV is equivalent
to that using the EFT approach.
The current best limit is $\epsilon\lesssim 0.0028$ for $M_{Z'}=$ 2 TeV
with combination of the three measurements.
We expect reducing the limit to 0.0025 with 300 fb$^{-1}$ data at
the LHC.
We note that all limits presented so far are for the choice of
$Z'$ width $\Gamma_{Z'}/M_{Z'}\equiv r$=0.1.
As mentioned earlier the constraints on $\epsilon$ scale as $\sqrt r$
approximately.
Thus for $r=0.05$ the current best limit would be $\epsilon\lesssim 0.0020$. 
We also calculate the projections for HL-LHC with a luminosity
of 3000 fb$^{-1}$, and only find improvements of a few percents for
the limits on NSIs in the full range of $M_{Z'}$ considered.
However, in future measurements with high statistics, one can
further extend the measured $p_T^{miss}$ to even higher values.
That requires a dedicated study on the SM backgrounds in that
region, and we expect more improvements can be gained than
those shown in Table.~\ref{Tab:HLLHC}.

\begin{table}[htpb]
  \centering
  \begin{tabular}{c|ccc|ccc|ccc}
  \hline
  ~ & \multicolumn{3}{c}{Current data} & \multicolumn{3}{c}{HL-LHC (300 fb$^{-1}$)} & \multicolumn{3}{c}{HL-LHC (3000 fb$^{-1}$)}\\
  \hline
  $M_{Z'}$ (TeV) & 0.2 & 2 & 100 & 0.2 & 2 & 100 & 0.2 & 2 & 100\\
  \hline
  $CMS_{J}$ & 0.039 & 0.0031 & 0.011 & 0.035 & 0.0030 & 0.010 & 0.035 & 0.0028 & 0.0097 \\
  $CMS_{W+Z}$ & 0.16 & 0.0043 & 0.014 & 0.16 & 0.0030 & 0.0093 & 0.15 & 0.0028 & 0.0088 \\
  $ATLAS_{J}$ & 0.078 & 0.0056 & 0.020 & 0.057 & 0.0052 & 0.018 & 0.057 & 0.0052 & 0.018 \\
  Combined & 0.042 & 0.0028 & 0.010 & 0.035 & 0.0025 & 0.0081 & 0.033 & 0.0023 & 0.0077 \\
 \hline
  \end{tabular}
  \caption{
	        Summary of current and projected 95\% CLs upper limit on NSIs 
		in a simplified $Z'$ model with $Z'$ mass of 0.2, 2, and 100 TeV respectively.
	We assume $ \epsilon_{u} = \epsilon_{d}=\epsilon$
	and $\Gamma_{Z^{\prime}}/M_{Z^{\prime}}=0.1$, and use
	the NLO predictions with theoretical uncertainties.
  \label{Tab:HLLHC}
	  }
\end{table}

LHC constraints on NC NSIs based on missing transverse energy have
also been studied in previous works~\cite{Choudhury:2018azm,Babu:2020nna}. 
It is noted that the data set taken by both works are
ATLAS mono-jet production with 36.1 fb$^{-1}$, and therefore partly overlapped
with this paper.
Our results concerning ATLAS mono-jet production are consistent with~\cite{Babu:2020nna},
however, less stringent than~\cite{Choudhury:2018azm}. 
By utilizing the more precise data from CMS mono-jet production with 35.9 fb$^{-1}$ data,
we have set a limit stronger by a factor of two than~\cite{Babu:2020nna},
which can be read from Table.~\ref{Tab:HLLHC}.
Projections for future run of LHC with integrated luminosities of 300 fb$^{-1}$
and 3000 fb$^{-1}$ have also been given in previous studies.
With consideration of reduced systematic uncertainties, they
expect larger improvement than in our study of which rather conservative
assumptions are taken.
On the other hand, there exist indirect searches of NSIs at the LHC
utilizing SM gauge symmetries.
Under consideration of the common $U(1)'$ coupling shared by
the $SU(2)_L$ doublet, process $pp \rightarrow Z' \rightarrow l^{+}l^{-} + X$
is taken into account, and measurements on dilepton
final state are used to set limits on NC NSIs~\cite{Han:2019zkz}.
Due to the better sensitivities for final state with charged leptons,
a stringent limit has been obtained on the common coupling $g$
between $Z'$ and fermions, to be about $10^{-2}$ for $M_{Z'}\approx 1$TeV,
based on ATLAS searches of dilepton resonances~\cite{Aad:2019fac}.
This limit can be converted into a limit on conventional coupling strength
of NSIs, $\epsilon \lesssim 10^{-5}$ through Eq.~(\ref{conv}).
Finally, it is worth noting that LHC constraints are fairly loose for light
mediators with mass smaller than electro-weak scale. 
For $M_{Z'}\approx$50 MeV, a limit of $\epsilon \lesssim 0.1$ has been
reached with the COHERENT experiment~\cite{Akimov:2018vzs,Han:2019zkz}.

\section{Conclusion}\label{sec:conc}

The study on possible non-standard interactions of neutrinos with
matter has a long history, and stringent limits have been imposed
from various experiments. 
The NSIs can affect the production, propagation as well as detection
of neutrinos, and have a direct consequence on global analysis of
neutrino properties like mass ordering and CP phases.
The successful operation of LHC opens new opportunities on searching
for neutrino NSIs at high momentum transfers which are complementary
to other experiments.
Neutrinos appear as signal of missing transverse momentums in detectors
same as those from dark matters.
There have been several studies on constraining NC NSIs
using measurements of mono-jet production at the LHC~\cite{Friedland:2011za,Choudhury:2018azm,Babu:2020nna}.
In our study we select various data sets from LHC measurements
at 13 TeV with integrated luminosities of $35\sim 139$ fb$^{-1}$,
including production of a single jet, photon, $W/Z$ boson, or charged lepton,
accompanied with large missing transverse momentums.
We derive constraints on neutral-current NSIs with quarks imposed by different
data sets in a framework of either effective operators or simplified
$Z'$ models.

We found the CMS measurement on mono-jet production gives the
strongest constraints on NSIs followed by the CMS measurement
on mono-$W/Z$ production.
The ATLAS mono-jet measurement also leads to comparable constraints
while the mono-photon and mono-lepton production show less sensitivities.
We use theoretical predictions of various production induced by NSIs calculated
at next-to-leading in QCD matched with parton showering and hadronizations.
The inclusion of higher order QCD effects stabilize the theory predictions
and result in more robust constraints.
In the framework of effective operators we find a 95\% CLs upper limit
of 0.010 on the conventional NSI strength parameter $\epsilon$.
In a simplified $Z'$ model we obtain
an upper limit on $\epsilon$ of 0.042 and 0.0028 for a $Z'$ mass
of 0.2 and 2 TeV respectively, assuming $\Gamma_{Z'}/M_{Z'}=0.1$.
Moreover, we discuss possible improvements from future runs of LHC
with higher luminosities.
We find a moderate reduction of the limits if using the same
experimental setups but expect further gains by extending current
measured missing transverse momentums to higher values.

\begin{acknowledgments}
This work was sponsored by the National Natural
Science Foundation of China under the Grant No. 11875189 and No.11835005,
and by the MOE Key Lab for Particle Physics, Astrophysics and Cosmology.
\end{acknowledgments}

\bibliography{lhcnsi}
\bibliographystyle{jhep}

\end{document}